\documentclass[reprint,aps,prd]{revtex4-1}
\usepackage{amsmath}
\allowdisplaybreaks[4]
\usepackage{multirow}
\usepackage{amsfonts}
\usepackage{xcolor}
\usepackage{comment}
\usepackage{graphicx}
\usepackage{amsmath}
\usepackage{siunitx}
\usepackage{float}
\usepackage[
    pdfauthor={Teng Zhang},
    pdftitle={Squeezing for heterodyne readout},
    colorlinks=true,
    linkcolor=black,
    urlcolor=blue,
    citecolor=blue
]{hyperref}

\DeclareSIUnit{\sqrthz}{\ensuremath{\sqrt{\text{\hertz}}}}

\begin{document}
\renewcommand\texteuro{FIXME}

\allowdisplaybreaks[4]
\title{Quantum Squeezing Schemes for Heterodyne Readout}

\author{Teng Zhang}
\email{tzhang@star.sr.bham.ac.uk}
\author{Denis Martynov}
\email{dmartynov@star.sr.bham.ac.uk}
\author{Andreas Freise}
\email{adf@star.sr.bham.ac.uk}
\author{Haixing Miao}
\email{haixing@star.sr.bham.ac.uk}
\affiliation{School of Physics and Astronomy, and Institute of Gravitational Wave Astronomy, University of Birmingham, Edgbaston, Birmingham B15\,2TT, United Kingdom}

\begin{abstract}

Advanced gravitational-wave detectors are limited by quantum noise in their most sensitive frequency band. Quantum noise suppression techniques, such as the application of the quantum squeezed state of light, have been actively studied in the context of homodyne readouts. In this paper, we consider quantum squeezing schemes for the heterodyne readouts. This is motivated by a successful suppression of the higher-order-mode content by stable recycling cavities in advanced detectors. The heterodyne readout scheme requires precise tuning of the interferometer parameters and a broadband squeezing source, but is conceptually simple and elegant. We further show that it is compatible with the frequency-dependent squeezing, which reduces both the shot noise and the radiation-pressure noise. We propose a test of the heterodyne readout with squeezing in Advanced LIGO. This can serve as a pathfinder not only for the implementation in future detectors, such as Einstein Telescope and Cosmic Explorer, but also for general
high-precision optical measurements. 

\end{abstract}
\maketitle

\section{Introduction}
\label{sec:introduction}
The Pound-Drever-Hall heterodyne technique~\cite{DreverPDH, Black_PDH_2001,Takahashi_2004,Sigg_2008,LIGO2015,Acernese_2006,Acernese_2014,Dooley_2015,PhysRevA.57.3898} is a powerful tool for stabilisation of optical cavities in modern precision instruments, such as frequency references for optical atomic clocks, passive laser gyroscopes, and gravitational-wave detectors.  In the heterodyne readout, phase modulated light probes the motion of the optical cavity and produces the signal on photodetectors at radio frequencies (RF).  
After demodulation, the residual signal is proportional to the cavity motion or the laser frequency noise. The heterodyne technique circumvents laser technical
noise by upconverting the signal detection to frequencies where the laser light is shot-noise-limited (a few MHz)~\cite{Rakhmanov:01} but couples technical noises of the modulation oscillator~\cite{Ward_THESIS_2010}.
In this paper, we do not focus on the technical noises of the heterodyne readout scheme and only consider the fundamental quantum noise.

Quantum noise in the heterodyne readout techniques has been previously studied in Refs.~\cite{PhysRevA.43.5022, PhysRevD.67.122005} in the context of gravitational-wave detectors. As it turns out, there are additional vacuum noises at twice the RF modulation frequency $2\omega_m$ away from the carrier frequency $\omega_0$, which leads to 50\% higher shot noise compared to the homodyne readout. The squeezed state of light can be used to suppress this additional noise, and 
in general, we need to have squeezing near the carrier frequency 
$\omega_0$ and the two RF sidebands frequencies 
$\omega_0\pm 2\omega_m$\,\cite{PhysRevA.44.4693,
PhysRevD.23.1693, PhysRevA.57.3898}.
In this paper, we further advance these studies and consider application of quantum noise suppression techniques to the heterodyne readout in advanced gravitational-wave detectors, \textit{i.e.}, a single squeezer is not only sufficient to improve the shot noise\,\cite{Gea-Banacloche1987}, 
but also can be made compatible with frequency-dependent squeezing, which reduces the quantum radiation-pressure noise. As discussed in Sec~\ref{sec:Sqz} and~\ref{sec:FD}, we find that quantum squeezing works for the heterodyne readout if (i) the source of squeezed states of light has a bandwidth at least twice the RF modulation frequency ($\omega_m$), (ii) the filter cavity for the frequency-dependent squeezing is tuned the same as in the current homodyne readout scheme, and (iii) the power imbalance of the upper and lower RF sidebands which are on phase quadrature is less than 10~\% for 12\,dB broadband squeezing. 
\begin{figure}[b]
\centering
  \includegraphics[width=1\columnwidth]{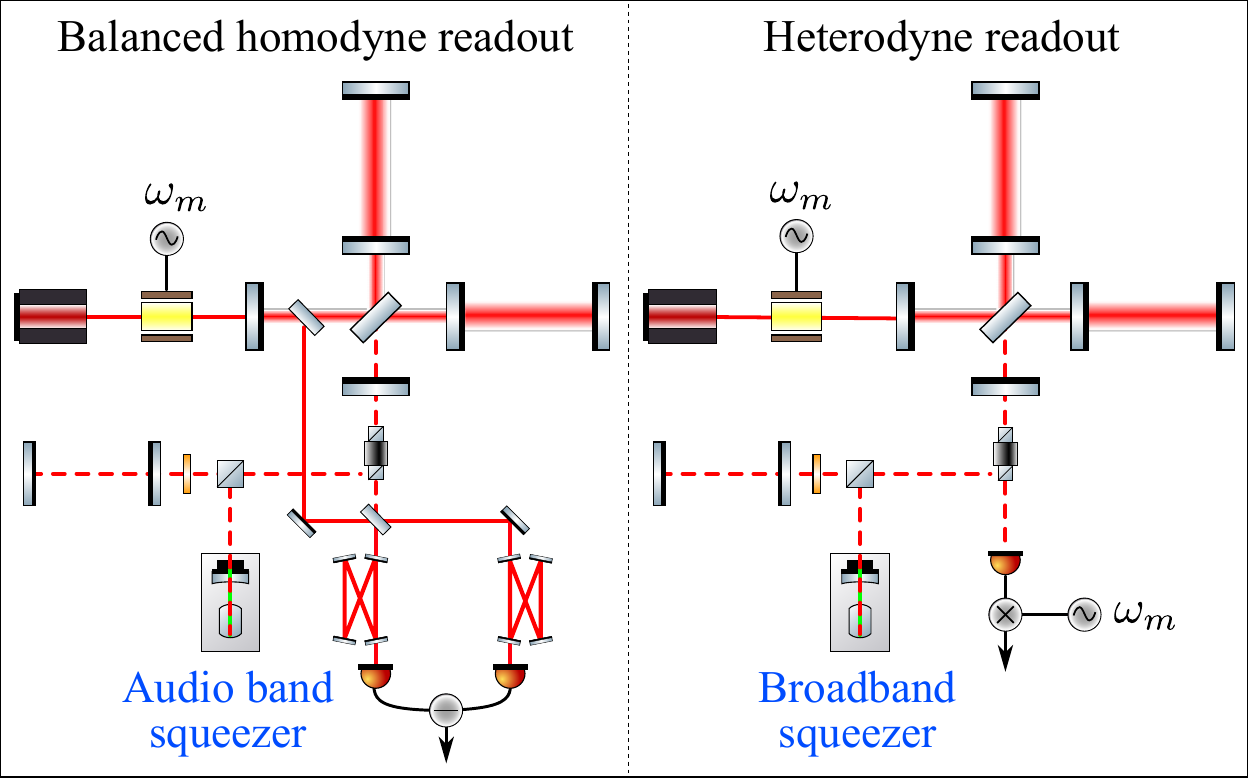}
\caption{Schematics of the balanced homodyne readout
and the heterodyne readout which requires a broadband
squeezer with bandwidth at least twice the 
modulation frequency $\omega_m$. } 
\label{fig:GWD}
\end{figure}

The main advantage of the homodyne readout scheme is that its quantum shot noise is a factor of $\sqrt{1.5}$ smaller compared to the one in the heterodyne readout for the same level of squeezing and optical losses. The price paid for this improvement is complexity involved in the balanced homodyne readout scheme which requires an additional local oscillator field and optical cavities to filter the carrier field from the RF sidebands\,\cite{Fritschel2014,PhysRevD.95.062001, Sebastian2015, Zhang_2018}. In Advanced LIGO, an additional local oscillator field is derived by offsetting the interferometer from its operating point. This scheme, known as DC readout~\cite{2015, Hild_2009, Fricke_2012, Grote_2010}, is also the current readout scheme in Advanced Virgo\,\cite{Acernese_2014}, KAGRA\,\cite{Akutsu:2019aa}, and  GEO\,600\,\cite{Hild_2009}. It has been very successful and allowed direct observation of gravitational waves for the first time~\cite{GW150914, GW170817, PhysRevX.9.031040}. However, the offset couples technical noise sources and will not allow future gravitational-wave detectors to reach their design sensitivity at low frequencies~\cite{PhysRevLett.120.141102}.

The heterodyne technique is ideal for coupled optical resonators and is already used in the Advanced LIGO detectors to stabilise auxiliary degrees of freedom and for initial stabilisation of the gravitational-wave channel. In this paper, we explore optical parameters when quantum noise in heterodyne and homodyne readout schemes are identical but heterodyne readout is conceptually simpler as shown in Fig.~\ref{fig:GWD}. This study is motivated by successful operation of the stable recycling cavities~\cite{Arain_RECYCLING_2008} in the Advanced LIGO detectors which has significantly suppressed higher-order-mode (HOM) content in the RF sidebands due to misalignments in the cavity~\cite{Gretarsson:07} and is also good for suppressing the HOM content from contrast defect. In Sec~\ref{sec:aLIGO_test} we show quantum noise of the Advanced LIGO detectors with the heterodyne readout and discuss steps for further improvements.

\section{Squeezing for Heterodyne readout}
\label{sec:Sqz}

\begin{figure}[t]
\centering
  \includegraphics[width=1\columnwidth]{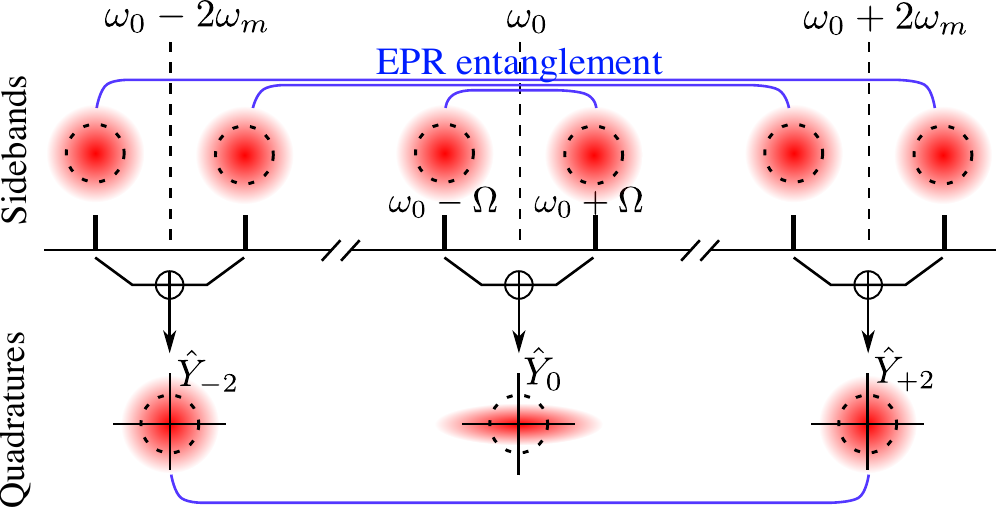}
\caption{Schematics of a broadband squeezer in the sideband  and the quadrature picture. In contrast for an audio band squeezer, the sidebands are entangled only for $\Omega$ up to kHz.} 
\label{fig:bd_sqz}
\end{figure}

In this section, we will show the resulting
quantum noise level for the heterodyne readout with a broadband squeezing. We choose the
demodulation phase such that the 
photocurrent is proportional to the phase 
quadrature $\hat Y$, which is given by: 
\begin{equation}\label{eq:HRp}
\hat Y(\Omega)={\hat Y_0}(\Omega)+\frac{{\hat Y_{+2}}(\Omega)+\xi \,{\hat Y_{-2}}(\Omega)}{1+\xi}\,,
\end{equation}
where $\xi$ denotes the amplitude ratio of two RF
sidebands which are on phase quadrature (more details are presented in Appendix\,\ref{sec:AHR}). 
Here ${\hat Y_0}$ and ${\hat Y_{\pm2}}$ represent the phase quadrature of three modes around the carrier frequency $\omega_0$ and the RF sidebands frequencies $\omega_0\pm 2\omega_m$, which are linear combinations 
of their audio sidebands, as illustrated 
in Fig.\,\ref{fig:bd_sqz}.  The last 
two terms contribute to the additional noise of
the heterodyne readout compared to the homodyne 
readout. The quantum noise level is quantified by 
the single-sided spectral density, which is defined  
for any operators $\hat A(\Omega)$ and $\hat B(\Omega')$: $
    \langle \psi |[\hat A \hat B^{\dag} +\hat B^{\dag}\hat A]/2|\psi \rangle \equiv \pi\, S_{AB}(\Omega) \delta(\Omega-\Omega')$. 
In the absence of squeezing, the quantum state $|\psi\rangle$
is in the vacuum state $|0\rangle$. 
Using the fact that $\langle 0| [\hat Y_{i}(\Omega)\hat Y^{\dag}_{j}(\Omega')+\hat Y^{\dag}_{j}(\Omega')\hat Y_{i}(\Omega)]/2|0\rangle = \pi\delta(\Omega-\Omega')\delta_{ij}$
$(i, j =0, \pm2)$, we have 
\begin{equation} 
    S_{YY} =1+\frac{1+\xi^2}{(1+\xi)^2}\,. 
\end{equation}
Considering the balanced case with $\xi=1$, \begin{equation}\label{eq:vac_balanced}
    S_{YY}|_{\rm balanced}=\frac{3}{2}\,, 
\end{equation}
which is 50\% higher than that of the homodyne readout. 

\subsection{ Audio band versus broadband squeezing}\label{sec:BS}

With the introduction of squeezing, the quantum noise level will be different, depending on the squeezing bandwidth. For the audio band squeezing, the squeezing is limited to the audio frequencies. The audio sidebands around the carrier frequency are entangled, which can be mapped to the (anti-)squeezing of the corresponding amplitude 
quadrature $\hat X_0$ and the phase quadrature $\hat Y_0$\,\cite{SCHNABEL20171}. 
For the phase squeezing, the spectral densities of the
quadratures satisfy the 
following covariance matrix,
\begin{equation}
\mathbb{V}_{o}=
\begin{bmatrix}
e^{2r_0}&0\\0&e^{-2r_0}
\end{bmatrix}\,,
\end{equation}
where $r_0$ is the squeezing factor at the audio frequencies. 
However, the sidebands around frequencies $\omega_o\pm 2 \omega_m$ are still in the vacuum state and are uncorrelated. Therefore, the audio band squeezing results in the spectral density for $\hat Y$: 
\begin{equation}\label{eq:SYYnarrow}
S_{YY}=e^{-2r_0}+\frac{1+\xi^2}{(1+\xi)^2}\,.
\end{equation}

\begin{figure}[t]
\centering
  \includegraphics[width=1\columnwidth]{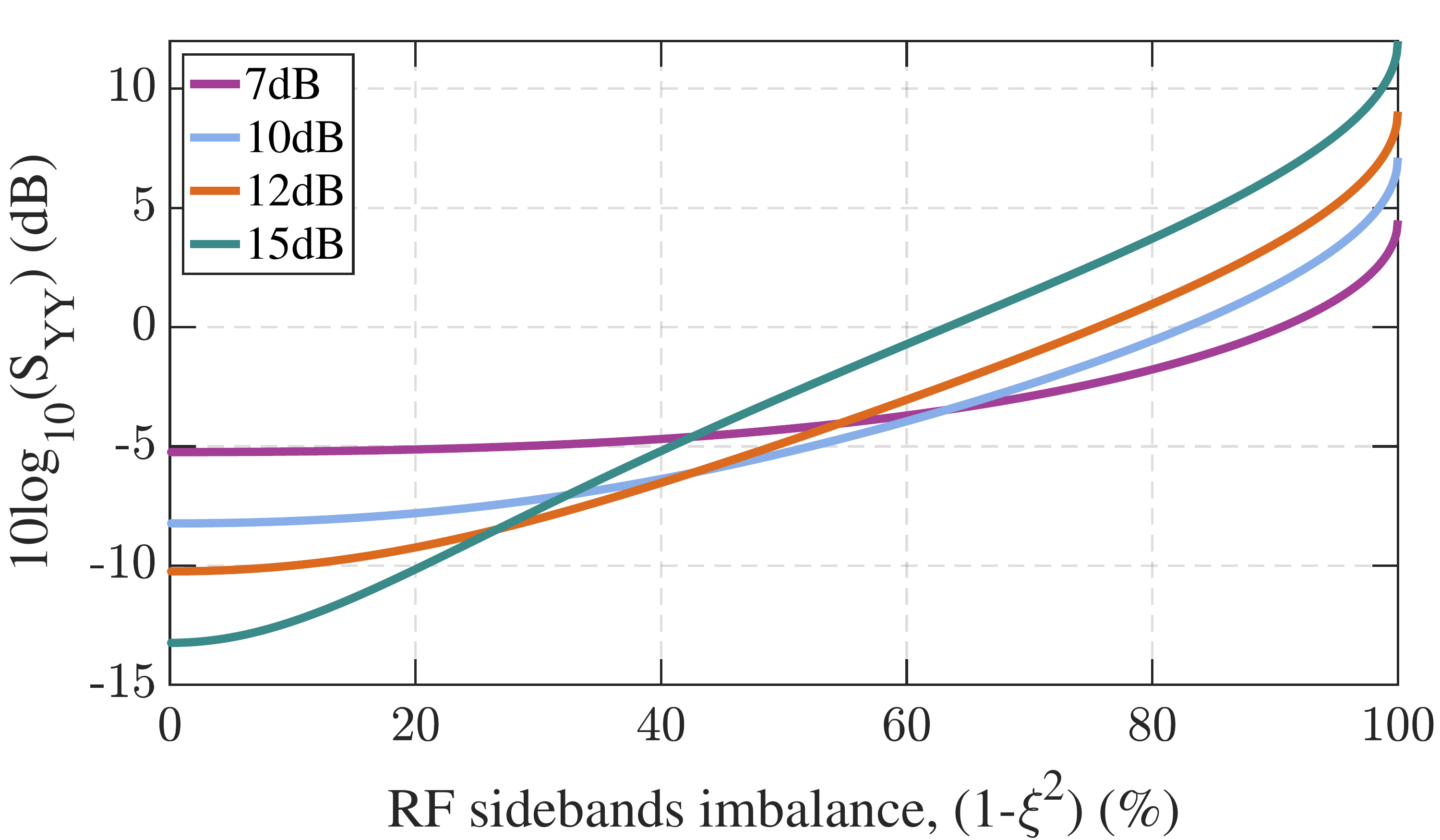}
\caption{Quantum noise level in dB for the heterodyne readout as a function of 
the power imbalance of two RF sidebands, considering different broadband squeezing lever.} 
\label{fig:imba}
\end{figure}

In contrast, as illustrated in Fig.~\ref{fig:bd_sqz}, for the broadband squeezing with a bandwidth up to RF, the sidebands around frequencies $\omega_0+2\omega_m$ and $\omega_0-2\omega_m$ are
entangled. Their corresponding quadratures also form the Einstein–Podolsky–Rosen (EPR) entanglement\,\cite{Gea-Banacloche1987, PhysRevA.67.054302, Ma:2017aa, SCHNABEL20171, Danilishin2019}, and for the phase squeezing, their spectral 
densities satisfy the following $4\times4$ covariance matrix: 
\begin{equation}\label{eq:Vepr}
\mathbb{V}_{\pm} = \begin{bmatrix}
\alpha &0 & \beta & 0\\
0 &\alpha &0& -\beta \\
\beta&0 &\alpha & 0\\
0& -\beta &0 &\alpha
\end{bmatrix}\,,
\end{equation}
where $\alpha=\cosh2 r_{2\omega_m}, \beta=\sinh 2r_{2\omega_m}$ with $r_{2\omega_m}$ denoting 
the squeezing factor at twice of the RF. As we can see, the uncertainties of the 
individual quadratures are larger than that of the vacuum, namely $S_{Y_{+2}Y_{+2}}=S_{Y_{-2}Y_{-2}}=\alpha\ge1$. However,
the sum of their phase quadratures, $\hat Y_{s} = (\hat Y_{+2}+\hat Y_{-2})/\sqrt{2}$, has uncertainty less than 1, namely, 
$S_{Y_s Y_s}={\alpha-\beta}=e^{-2r_{2\omega_m}}$. With a broadband squeezer, the spectral density for $Y$ reads
\begin{equation}
S_{YY}=\frac{3}{2}e^{-2r}+\left(\frac{1-\xi}{1+\xi}\right)^2\frac{e^{2r}}{2}\,,
\end{equation}
where we have assumed $r_0=r_{2\omega_m}\equiv r$ for 
simplicity.
If two RF sidebands were balanced with $\xi=1$,
\begin{equation}\label{eq:SqzSyy}
S_{YY}|_{\rm balanced}=\frac{3}{2}e^{-2r}\,.
\end{equation} 
The additional noise due to the fluctuations around $\omega_0\pm2\omega_m$ is smaller by a factor of $e^{2r}$ comparing with Eq.\,\eqref{eq:vac_balanced}. 
Fig.~\ref{fig:imba} shows $S_{YY}$ as a function of the sidebands imbalance. With 10\,\% imbalance in the sideband power, \textit{i.e.}  
$\xi = \sqrt{0.9}$, there is still around 10\,dB quantum noise suppression for 12\,dB input broadband squeezing.

\subsection{Frequency-dependent squeezing}
Frequency-dependent squeezing 
has been proposed to simultaneously suppress 
the shot noise 
and the quantum radiation-pressure noise in gravitational-wave detectors\,\cite{kimble2001}. It uses a 
Fabry-Perot cavity as the filter cavity to transform the squeezed light, and will be 
implemented in, e.g., the Advanced LIGO plus upgrade. 
In this section, we will show the suppression of the additional noise with the broadband squeezing also holds for the frequency-dependent squeezing. 

\begin{figure}[t!]
\centering
  \includegraphics[width=1\columnwidth]{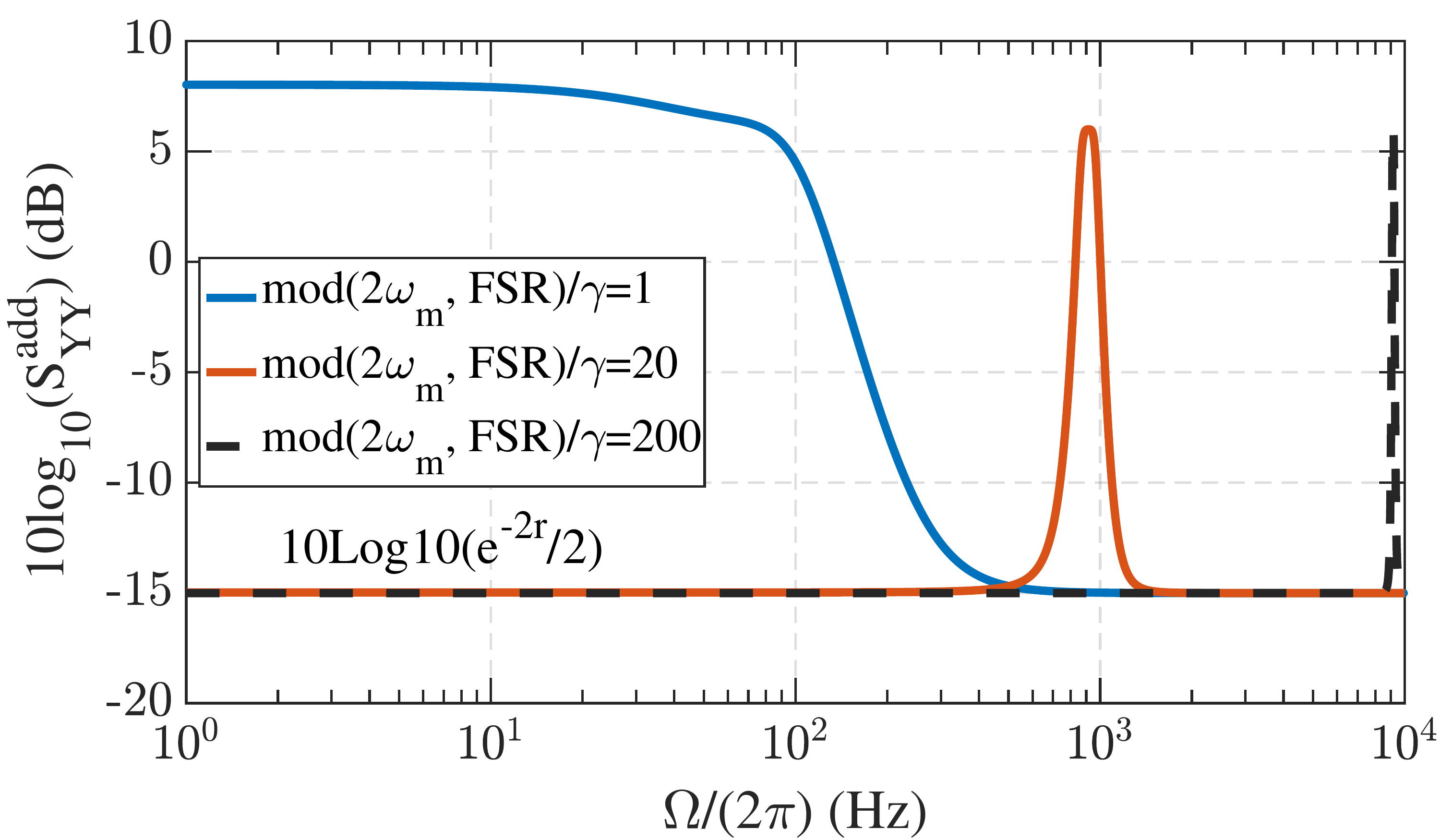}
\caption{Plot shows the level of the additional noise from $\omega_0\pm 2\omega_m$
in the phase quadrature at the reflection port of the filter cavity for different RF $\omega_m$. 
The broadband squeezing level is assumed to be 12\,dB. The optical parameters of the filter cavity are the same as A+ design. }
\label{fig:saddF}
\end{figure}

The filter cavity imprints different 
phases on the sidebands, which effectively 
creates a frequency-dependent rotation of the quadratures. Mathematically, the rotation is described by the following transfer matrix\,\cite{SD2012}: 
\begin{equation}
\mathbb{R}=e^{i\Phi}\begin{bmatrix}
\cos \theta & -\sin \theta \\\sin \theta &\cos \theta
\end{bmatrix}\,.
\end{equation}
Here the phase $\Phi$ and rotation angle $\theta$ are 
\begin{equation}
\Phi=\rm{atan} \frac{2\gamma\Omega}{\gamma^2+\Delta^2-\Omega^2}\,,  \theta=\rm{atan} \frac{2\gamma\Delta}{\gamma^2-\Delta^2+\Omega^2}\,.
\end{equation}
The frequency $\gamma$ is the filter cavity bandwidth. 
The frequency $\Delta$ is the cavity detuning, and is
different for the three 
modes: $\Delta \equiv \Delta_0$ for 
the mode around the carrier frequency; for RF modes around $\omega_0\pm 2\omega_m$, 
\begin{equation}
\Delta_{\pm2}\equiv \Delta_0\pm\rm{mod}(2\omega_{m}, \rm{FSR})\,,
\end{equation}
where FSR is the free spectral 
range of the filter cavity. 

In the balanced case, the spectral density of the additional noise due to fluctuations around $\omega_0\pm 2\omega_m$ is 
\begin{equation}
S_{Y Y}^{\rm add}=\frac{1}{2}\left[\alpha-\beta\cos(\Phi_{+2}-\Phi_{-2})\cos(\theta_{+2}+\theta_{-2})\right]\,, 
\end{equation}
where 
$\Phi_{\pm 2}$ and $\theta_{\pm 2}$ are the phase and rotation angle for quadratures of $\omega_0\pm 2\omega_m$. Ideally, we want 
$\Phi_{\pm2} = \theta_{\pm2} = 0$, so that $S_{YY}^{\rm add} = (\alpha-\beta)/2 = e^{-2r_{2\omega_m}}/2$, which 
leads to the minimum 
additional noise. This can 
be approximately achieved
when $\Delta_{\pm 2}$ 
is much larger than 
the filter cavity 
bandwidth $\gamma$, namely 
having $2\omega_m$ away 
from any FSR of the filter
cavity. 
As an illustration, in Fig.~\ref{fig:saddF}, 
we show the additional 
noise as a function 
of the distance of 
$2\omega_m$ away 
from the FSR (normalised 
by $\gamma$). We assume 
a filter cavity parameter the same as the A+ design, namely,
the cavity bandwidth  $\gamma/(2\pi) = \Delta_0/(2\pi)=45.8\,{\rm Hz}$. Indeed, when $2\omega_m$ is offset from $N\times {\rm FSR}$ by 200 times of the filter cavity bandwidth, the additional noise level is close to $e^{-2r_{2\omega_m}}/2$ for the entire frequency band relevant to gravitational-wave signals. Here $N$ is an arbitrary integer.

\begin{figure}[t]
\centering
\includegraphics[width=1\columnwidth]{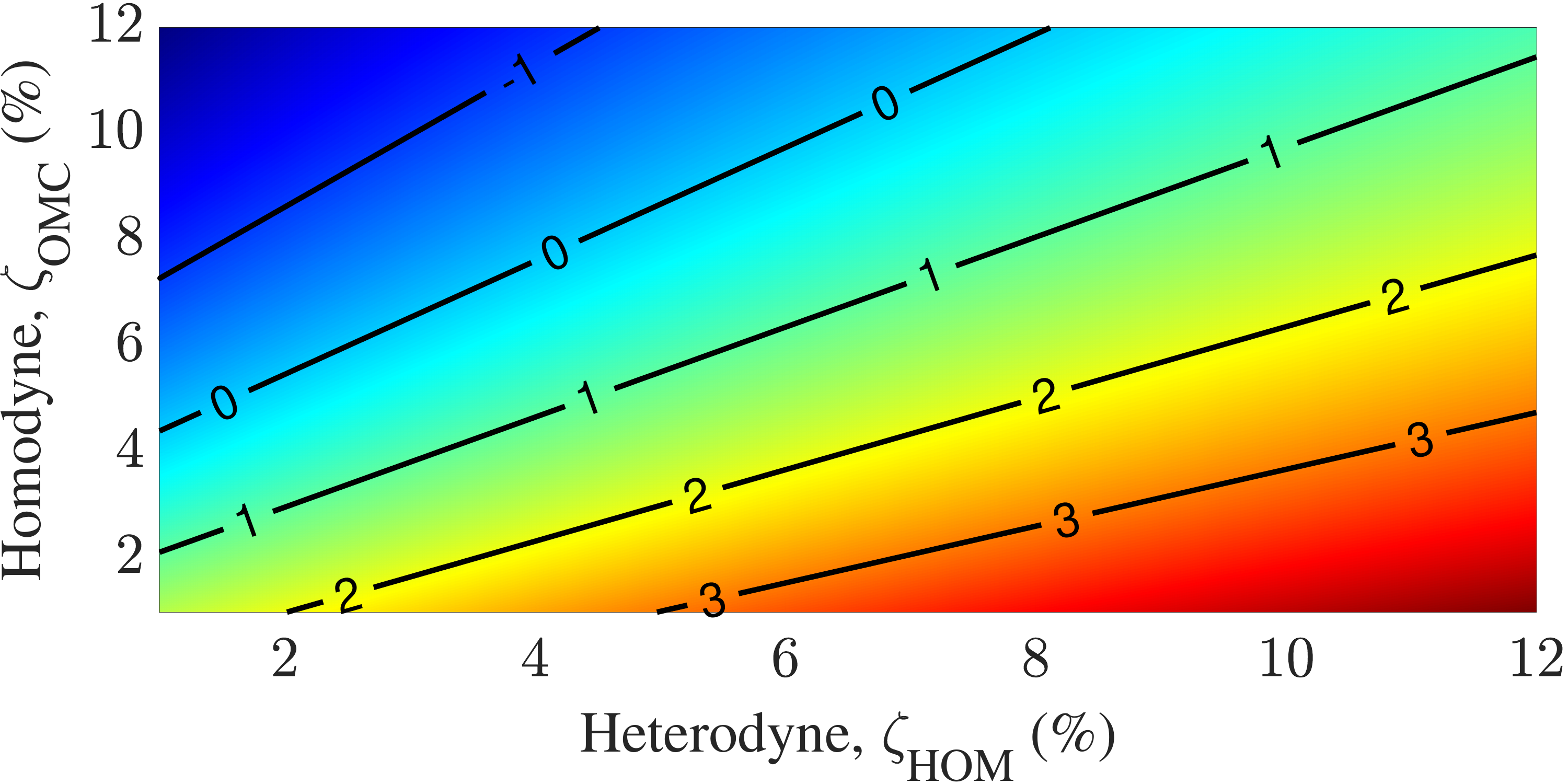}
\caption{ Figure shows the ratio of 
the quantum noise spectral density of the heterodyne readout
over the that of homodyne readout 
as a function of the OMC loss ($\zeta_{\rm OMC}$) in homodyne readout and HOM content ($\zeta_{\rm HOM}$) in heterodyne readout in the unit of dB. The contour line of $0$\,dB denotes the cases when the two noise levels are equal. 12\,dB input squeezing is assumed.} 
\label{fig:HM}
\end{figure}

\section{Higher-order-modes
and Schnupp asymmetry}
\label{sec:FD}
In comparison with the homodyne readout, 
there are two new 
quantum-noise related issues associated 
with the heterodyne readout. The first issue is the 
higher-order-mode content leaking through the dark port
due to the absence of output mode cleaner, which 
shall be traded off with the benefit of 
removing the output mode cleaner that induces the mode mismatch and misalignment loss. The second issue has to do with vacuum fluctuations around $\omega_0\pm 2\omega_m$ which transmit from the bright port to the dark port due to the Schnupp asymmetry. They act like additional optical losses. 

In the absence of output mode cleaner, the high-order-mode content at the carrier frequency and in the RF sidebands will both introduce additional quantum noise at $\omega_0\pm\omega_m$, which are in the vacuum sate. We define the ratio of the power of the higher-order-mode content  to the total sideband power as $\zeta_{\rm HOM}=\zeta_{\rm HOM}^{0}+\zeta_{\rm HOM}^{\omega_m}$, in which $\zeta_{\rm HOM}^{0}$ represents carrier frequency components and $\zeta_{\rm HOM}^{\omega_m}$ represents RF components.
In the case of balanced RF sidebands, the total quantum noise spectral density of the heterodyne readout is
\begin{equation}
S_{YY}^{\rm Heterodyne}=\frac{3}{2}e^{-2r}+\zeta_{\rm HOM}^{0}+m\zeta_{\rm HOM}^{\omega_m}\,,
\end{equation}
where $m$ is between 1 and 1.5 depending on the mode coherence between upper and lower RF sideband.  
For the homodyne readout, the output mode cleaner loss, quantified by
$\zeta_{\rm OMC}$, also leads to a degradation of the squeezing, namely, 
\begin{equation}
S_{YY}^{\rm Homodyne}=(1-\zeta_{\rm OMC})e^{-2r}+\zeta_{\rm OMC}\,.
\end{equation}
In Fig.\,\ref{fig:HM}, we show the ratio of these two spectral densities in dB as a function of $\zeta_{\rm HOM}$ and $\zeta_{\rm OMC}$. We take the the lower bound shot noise contribution from HOM content in RF sidebands, \textit{i.e.} $m=1$. 

\begin{figure}[t]
\centering
\includegraphics[width=1\columnwidth]{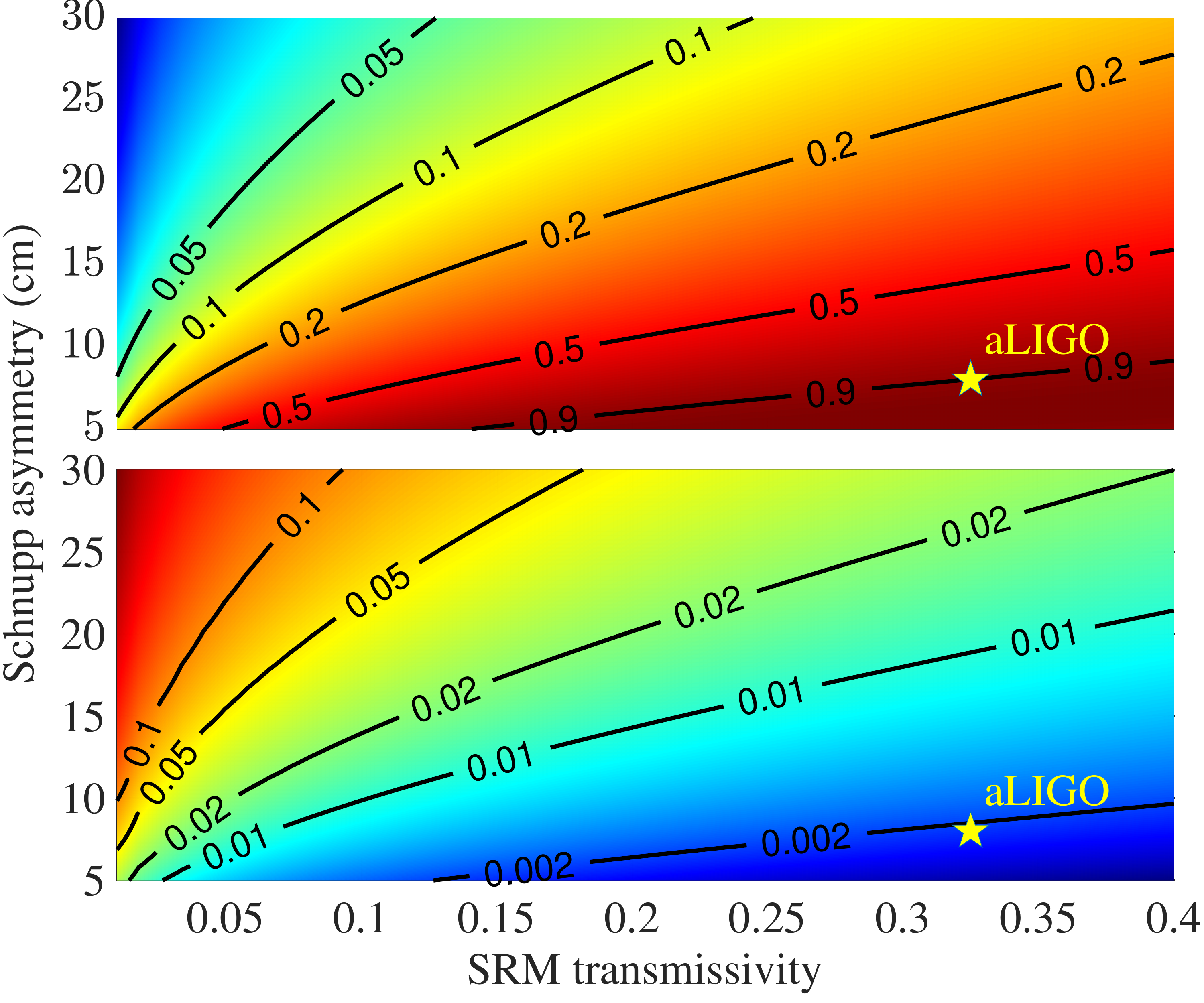}
\caption{The top panel
shows the transmissivity for 
45\,MHz sidebands. 
The bottom panel is the equivalent optical 
loss for squeezing around 90\,MHz 
as a function of the Schnupp 
asymmetry and the SRM transmissivity. 
The stars are rough 
estimates for the current situation of Advanced 
LIGO (aLIGO).} 
\label{fig:Schnnup}
\end{figure}

The Schnupp asymmetry allows the RF 
sidebands from the 
bright port to transmit to the readout 
port (dark port) as the local 
oscillator for the heterodyne readout. 
However, it also couples the vacuum noise 
from the bright port to the dark port, which is equivalent to introducing optical loss to the broadband squeezing near 
twice of the RF. Such a loss is determined by the transmissivity of the coupled power and signal recycling cavities, which depends on the optical properties of
three components: the power recycling mirror, the signal recycling mirror, and the central Michelson. In particular, the effective amplitude transmissivity and reflectivity of the central Michelson is, according to Ref.\,\cite{Izumi_2016}, $t_{\rm MI}=-\sin\omega_m\Delta L/c\,,r_{\rm MI}=-\cos \omega_m\Delta L/c$, where $\Delta L$ is the Schnupp asymmetry and $c$ is the speed of light. In advanced LIGO, the 45MHz sidebands resonate in both power recycling cavity and signal recycling cavity. The resonate condition in power recycling cavity builds on the accumulated phase $\pi+2N\pi$ of the 45MHz sidebands traveling through round macroscopic length of power recycling cavity and $\pi$ phase shift acquired from the arm cavity, in which the 45MHz RF sidebands are anti-resonance. In signal recycling cavity, the 45MHz sidebands accumulate phase $2N\pi$ traveling round trip of signal recycling cavity under \textit{Resonate Sidebands Extraction} mode.  The resonance condition of 90MHz filed is different, it still resonates in signal recycling cavity but anti-resonates in power recycling cavity.
The effective optical loss for 90\,MHz and the transmissivity for 45\,MHz are shown in the  Fig.~\ref{fig:Schnnup}. As we can see, in current aLIGO configuration, the optical loss for 90\,MHz squeezing fields is around $0.2\%$.
And the transmission of 45MHz sidebands can be adjusted significantly by tuning the Schnupp asymmetry or the signal recycling mirror transmissivity without boosting the optical loss at 90MHz significantly.

\section{Heterodyne in Advanced LIGO}
\label{sec:aLIGO_test}


In this section, we illustrate our findings in Sec.~\ref{sec:Sqz} and Sec.~\ref{sec:FD} on the example of the Advanced LIGO detectors which currently operate with squeezing and homodyne readout~\cite{Tse_SQZ_2019}. The source injects $7.2 \pm 0.3$\,dB of squeezing and the maximum observed level is $3.2 \pm 0.1$ dB. The discrepancy between the amount of injected and observed squeezing comes from the optical losses in the interferometer. According to Ref.~\cite{Tse_SQZ_2019}, around $25\%$ of the signal is lost on the Faraday isolators, on the output mode cleaner, and photodiodes. Another 10\,\% of the signal is lost on some optical components yet to be identified.

If switched to the heterodyne readout, the quantum noise level in the Advanced LIGO detectors would be 
\begin{equation}\label{eq:adv_ligo_shot}
\begin{split}
S_{YY}^{\rm total} =& (1-\epsilon_0)e^{-2r_0} + \epsilon_0+\\  &\frac{1}{2}\left[(1-\epsilon_{2\omega_m})e^{-2r_{2\omega_m}}+\epsilon_{2\omega_m})\right]+\zeta_{\rm HOM}^{0}+\zeta_{\rm HOM}^{\omega_m}\,,
\end{split}
\end{equation}
where $\epsilon_0$ and $\epsilon_{2\omega_m}$ are optical losses of the interferometer near the carrier frequency and RF sidebands at $2\omega_m$, respectively.
For the current Advanced LIGO configuration, according to Ref.~\cite{Araj_OMC_Scan_1988}, $\zeta_{\rm HOM}=0.12$ and the higher-order-mode content is dominated by the 02 modes of the RF sideband. The bandwidth of the LIGO squeezing source is $\sim$10\,MHz~\cite{Oelker_THESIS_2016}, which is much smaller than $2\omega_m = 90$\,MHz, and therefore, we have $r_{2\omega_m} \approx 0$. Furthermore, the loss from the mode matching and alignment of the output mode cleaner which is 5-15\,\% in the homodyne readout scheme is negligible in the heterodyne readout, as it does not require output mode cleaner. Using the parameters discussed above and Eq.(\ref{eq:adv_ligo_shot}) we estimate the final observed squeezing is around $1.7 \pm 0.2$\,dB for the heterdoyne readout in the current Advanced LIGO detectors.  

The LIGO parameters can be optimised for the heterodyne readout as discussed in Sec.~\ref{sec:Sqz} and Sec.~\ref{sec:FD}. First, we need to increase the bandwidth of the squeezing source above $\sim100$\,MHz. Achieving this milestone will make $r_{2\omega_m}=r_0$. The observed squeezing level will be increased up to  $3.9 \pm 0.2$\,dB. 
Finally, if we can suppress the 02 mode of the RF sidebands at the antisymmetric port, we will reduce the higher-order-mode content down to $\zeta_{\rm HOM}=0.02$, and the Advanced LIGO detectors will reach the observed level of squeezing equal to $4.7 \pm 0.3$\,dB. 

The corresponding shot noise amplitude spectral density is only a factor of $\approx 1.02$ larger than the current shot noise in the Advanced LIGO detectors, which is much smaller compared to a factor of $\sqrt{1.5}\approx1.22$ for the case of the same optical losses in both readouts.
Moreover, higher input squeezing will further shrink the gap between the homodyne and heterodyne readout in our model since we remove the output mode cleaner for the heterodyne readout. Performing a test described in this section has a strong potential to help the Advanced LIGO detectors to estimate losses and technical noises of the output mode cleaner. The test will also demonstrate that squeezing of the quantum noise in the heterodyne readout works according to our model and can be considered for future gravitational-wave detectors.

So far, we have been focused on the cases with RF sidebands having nearly equal power. The imbalance was treated as an imperfection, which is the case for gravitational-wave detectors, as the imbalance usually introduces undesired 
technical noises. If the technical noises can be suppressed, we can consider more general 
heterdyne readouts with strongly imbalanced sidebands, or even a single
sideband. However, a single broadband squeezing 
 will not be able to suppress the additional quantum noise from $\omega_0\pm 2\omega_m$, 
and we would need three-mode squeezing schemes shown in Appendix~\ref{sec:TM}. 

\section{Conclusion and discussion}\label{sec:conc}
To summarise, we have investigated squeezing schemes for the heterodyne readout in advanced gravitational-wave detectors. Our research shows that the heterodyne readout is compatible with frequency-dependent squeezing by using a broadband squeezer and a filter cavity the same as the one for the homodyne readout. We have studied the problem of the higher-order-mode content leaking to the dark port and the vacuum noise coupling from the bright port to the dark port around frequencies $\omega_0\pm 2\omega_m$ due to the Schnupp asymmetry, which turns out to be negligible. Taking Advanced LIGO for instance, there is a promising path to reduce the higher-order-mode content with the stable signal recycling cavity and the suppression of the dominant 02 mode. 
The heterodyne readout requires less auxiliary optics, and its sensitivity can be made comparable to that of the balanced homodyne readout. 

The strategies of incorporating quantum squeezing into general heterodyne readout can be applied to a broad class of optical measurements that use RF sidebands as in the Pound–Drever–Hall technique. Therefore, our findings will not only have impacts to the gravitational-wave community but also the general high-precision measurement community. 

\section{Acknowledgements}
We would like to thank Roman Schnabel, Ken Strain, Stefan Hild, Joseph Briggs, Lee McCuller and Daniel Sigg for fruitful discussions. T. Z., D. M., A. F. and H. M. acknowledge the support of the Institute for Gravitational Wave Astronomy at University of Birmingham. A. F. has been supported by a Royal Society Wolfson Fellowship which is jointly funded by the Royal Society and the Wolfson Foundation. H. M. is supported by UK STFC Ernest Rutherford Fellowship (Grant No. ST/M005844/11). 

\appendix
\section{Description of heterodyne readout}\label{sec:AHR}

In heterodyne readout, the RF sidebands are generated by modulating the phase of the carrier field with a RF sinusoidal signal $\left[m+\delta m (t)\right]\cos\left(\omega_m t+\delta \phi (t)\right)$,
where $m$ is the modulation index, $\omega_m$ is the modulation frequency, $\delta m (t)$ is the modulation index fluctuations and $\delta \phi (t)$ is the modulation phase fluctuations. When $m\ll1$, the laser field can be described approximately as
\begin{equation}\label{eq:E}
\begin{split}
E(t)=&E_0(1-\frac{m^2}{4})e^{i\omega_0t}+\\
&E_0\left[\frac{i m}{2}+\frac{i\delta m(t)}{2}\right]e^{i\omega_0t}\left(e^{-i\omega_mt}+e^{i\omega_mt}\right)+\\&E_0\frac{m\delta \phi(t)}{2}e^{i\omega_0t}\left(e^{-i\omega_mt}-e^{i\omega_mt}\right)+h.c.\,,
\end{split}
\end{equation}
where $E_0$ is the amplitude of the carrier field, $h.c.$ denotes the hermitian conjugate.
According to Eq.~\eqref{eq:E}, we symbolise RF local oscillator as 
\begin{equation}\label{eq:L}
\begin{split}
L(t)=&\left[L_{+}+ l_+(t)\right]e^{i(\omega_0+\omega_m)t}+\\&\left[L_{-}+ l_-(t)\right]e^{i(\omega_0-\omega_m)t}+h.c.\,,
\end{split}
\end{equation}
where $L_+,L_-$ are the upper and lower RF sidebands with $L_+=L_-=i\frac{m}{2}E_0$. $l_+,l_-$ are the fluctuation terms of the local oscillator beam. Their classical parts have the amplitude,
\begin{subequations}\label{eq:dl}
\begin{align}
l_+(t)=\left[\frac{i\delta m(t)-m\delta \phi(t) }{2}\right]E_0\,,
\\
l_-(t)=\left[\frac{i\delta m(t)+m\delta \phi(t) }{2}\right]E_0\,.
\end{align}
\end{subequations}
We define the signal field $Z(t)$ and specify only three modes around frequencies $\omega_0$ and $\omega_0\pm2\omega_m$, which will eventually contribute to the final output. There is 
\begin{equation}\label{eq:o}
\begin{split}
Z(t)=&\left[\overline{Z}_0+Z_0(t)\right]e^{i\omega_0t}+\\
&Z_+(t)e^{i(\omega_0+2\omega_m)t}
+Z_-(t)e^{i(\omega_0-2\omega_m)t}+\\&h.c.\,,
\end{split}
\end{equation}
where $\overline{Z}_0$ is the amplitude of the carrier that leaks to the dark port of the interferometer, $Z_0(t)$ represents the fluctuations around frequency $\omega_0$, including both signal and noise. $Z_+(t)$ and $Z_-(t)$ represent the fluctuations around frequencies $\pm \omega_m$, respectively. Ignoring the second order terms and the terms at irrelevant frequencies, the beat between local oscillator and signal field can be calculated as
\begin{equation}\label{eq:phc}
\begin{split}
\left[L(t)+Z(t)\right]\left[L(t)+Z(t)\right]^{\dagger}=\\
2\left[L_++l_+(t)\right]\left[\left(\overline{Z}_o+Z_o(t)\right)^{\dagger}e^{i\omega_m t}+Z_+^{\dagger}(t)e^{-i\omega_mt}\right]+\\2\left[L_-+l_-(t)\right]\left[\left(\overline{Z}_o+Z_o(t)\right)^{\dagger}e^{-i\omega_m t}+Z_-^{\dagger}(t)e^{i\omega_mt}\right]+\\
h.c.\,.
\end{split}
\end{equation}
This photocurrent is then demodulated by the sinusoidal signal $\left[m'+\delta m' (t)\right]\cos(\omega_m t+\phi'+\delta \phi' (t))$, which is from the same source of modulation signal. It can be written approximately as,
\begin{equation}\label{eq:m}
\left[m'+\delta m'(t)\right]\cos(\omega_m t+\phi')-\delta \phi'(t) m'\sin(\omega t+\phi')\,,
\end{equation}
where $m'$ and $\delta m'(t) $ represent the amplitude of demodulation signal and its fluctuations; $\phi'$ and $\delta \phi'(t)$ represent the demodulation phase and its fluctuations. After applying a low pass filter with audio bandwidth to the product of Eq.~\eqref{eq:phc} and ~\eqref{eq:m}, we can get the demodulated output. 

It is convenient to describe the output in frequency domain using quadrature operator base on the relation\,\cite{PhysRevD.67.122005}
\begin{equation}
A\hat{Z}^{\dagger}_{-\Omega}+A^{\dagger}\hat{Z}_{\Omega}=\sqrt{2}|A|\hat{Z}_{\zeta}(\Omega),
\end{equation}
where $A=|A|e^{i\zeta}$ is an arbitrary complex amplitude. The quadrature operator $Z_{\zeta}$ is defined as 
\begin{equation}\label{eq:b}
Z_{\zeta}(\Omega)=\hat{X}(\Omega)\cos\zeta+\hat{Y}(\Omega)\sin{\zeta}\,,
\end{equation}
with
\begin{equation}
\hat{X}(\Omega)=\frac{\hat{Z}_{\Omega}+\hat{Z}^{\dagger}_{-\Omega}}{\sqrt{2}}\,,Y(\Omega)=\frac{\hat{Z}_{\Omega}-\hat{Z}^{\dagger}_{-\Omega}}{\sqrt{2}i}\,,
\end{equation}
representing amplitude quadrature and phase quadrature, respectively.
Eventually, the demodulated output can then be calculated as
\begin{widetext}
\begin{equation}\label{eq:I}
\begin{split}
I(\Omega)&=
\sqrt{2}m'\left[|L_0|{Z_0}_{\zeta_0}(\Omega)+|L_+|{Z_+}_{\zeta_+}(\Omega)+|L_-|{Z_-}_{\zeta_-}(\Omega) \right]
+\sqrt{2}m'|\overline{Z}_0|\left[{l_+}_{\alpha_{+}}(\Omega)+ {l_-}_{\alpha_{-}}(\Omega)\right]\\
&-2m'\delta\phi'(\Omega)\left[|L_+||\overline{Z}_0|\cos\beta_++|L_-||\overline{Z}_0|\cos\beta_-\right]+2\delta m'(\Omega)\left[|L_+||\overline{Z}_0|\cos \psi_++|L_-||\overline{Z}_0|\cos \psi_-\right]\,,
\end{split}
\end{equation}
\end{widetext}
where
\begin{equation}
|L_0|={|L_+e^{-i\phi'} +L_-e^{i\phi'}|}\,,
\end{equation}
and
\begin{subequations}
\begin{align}
 \zeta_0&={\rm arg}\left(L_+e^{-i\phi'}+L_-e^{i\phi'}\right)\,, \\\zeta_\pm&=\pm\phi'+{\rm arg}L_\pm\,,\\
\alpha_{\pm}&=\pm \phi'+{\rm arg}\overline{Z}_{0}\,,\\
\beta_+&= {\rm arg}\left(L_+e^{-i\phi'}\right)-{\rm arg}\overline{Z}_0-\frac{\pi}{2}\,,\\
\beta_-&= {\rm arg}\left(L_-e^{i\phi'}\right)-{\rm arg}\overline{Z}_0+\frac{\pi}{2}\,,\\
\psi_+&= {\rm arg}\left(L_+e^{-i\phi'}\right)-{\rm arg}\overline{Z}_0\,,\\\psi_-&= {\rm arg}\left(L_-e^{i\phi'}\right)-{\rm arg}\overline{Z}_0\,.
\end{align}
\end{subequations}

When $\overline{Z}_0=0$, considering balanced RF sidebands which are on phase quadrature and $\phi'=0$ for phase measurement, there is
\begin{equation}
\begin{split}
I(\Omega,|L_+|=|L_-|)=
\\\sqrt{2}m'|L_0|\left[{Y_0}(\Omega)+\frac{{Y_+}_2(\Omega)+{Y_-}_2(\Omega)}{2}\right]\,,
\end{split}
\end{equation}
where $Y_0$ and $Y_{\pm2}$ represent the phase quadrature of three modes around frequency $\omega_0$ and $\omega_0\pm 2\omega_m$.
If, $|L_+|\ne|L_-|$, 
\begin{equation}
\begin{split}
I(\Omega,|\frac{|L_-|}{|L_+|}=\xi)=
\\\sqrt{2}(1+\xi)m'|L_+|\left[{Y_0}(\Omega)+\frac{{Y_+}_2(\Omega)+\xi{Y_-}_2(\Omega)}{1+\xi}\right]\,,
\end{split}
\end{equation}
where $\xi=|\frac{L_-}{L_+}|$ denotes the ration of amplitude of two RF sidebands on phase quadrature.
\section{Three modes squeezing schemes}\label{sec:TM}
\begin{figure}[t]
\centering
  \includegraphics[width=1\columnwidth]{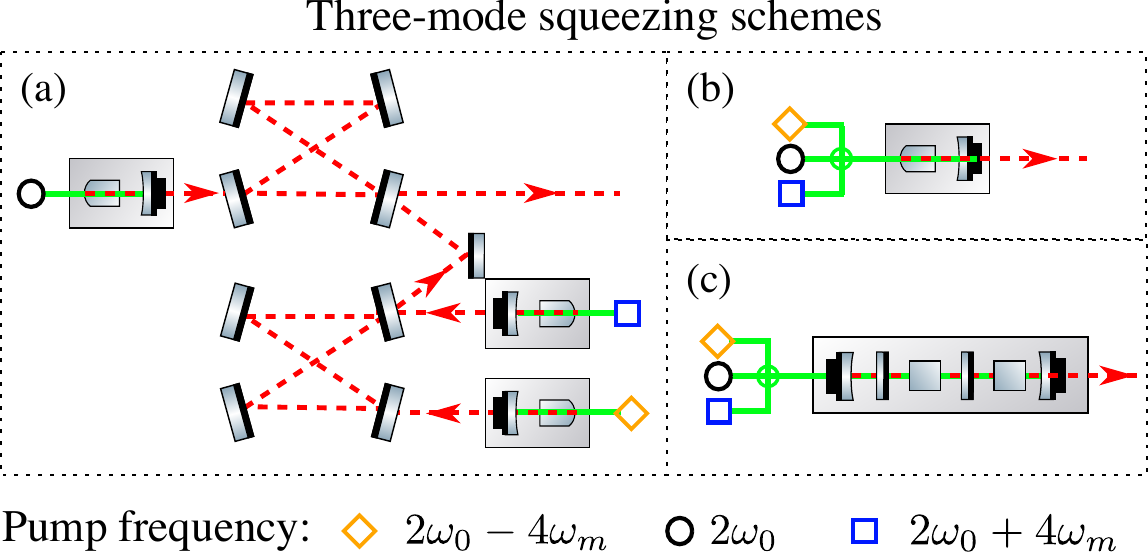}
\caption{Different three-mode squeezing schemes to produce independent squeezing fields near $\omega_0-2\omega_m$, $\omega_0$, and $\omega_0+2\omega_m$ for general imbalanced heterodyne readouts. 
Scheme (a) coherently combines the 
squeezing from three independent sources.
Scheme (b) uses three longitudinal modes of a single cavity, which are separated by the free spectral range equal to $2\omega_m$. It works when the crystal squeezing bandwidth is smaller than $2\omega_m$. Otherwise, a more sophisticated scheme (c) with three cavities coupled together can be an option. } 
\label{fig:3mode_sqz}
\end{figure}

In this appendix, we introduce squeezing schemes as shown in Fig.~\ref{fig:3mode_sqz}, which produce three independent squeezing modes around $\omega_0$ and $\omega_0\pm2\omega_m$. 

Scheme (a) uses three audio band squeezing sources which are coherently combined through two
optical cavities that properly reflect and transmit the fields using their frequency selectivity. Scheme (b) uses a single cavity
with three pumps interacting with the non-linear
crystal. The longitudinal mode frequency of the cavity coincide with that of three squeezing modes. The squeezing
bandwidth of the crystal, however, needs to be smaller than
$2\omega_m$ to avoid the EPR entanglement between
neighbouring modes. If it is challenging to achieve, we can consider scheme (c) which uses coupled cavities and two nonlinear crystals.  

In scheme (c), we define the optical modes of the three cavities as $a$, $b$ and $c$ with identical frequency $\omega_{E}$. The power transmissivities of the mirrors between cavities $a$, $b$ is defined as $T_1$; the power transmissivities of the mirrors between cavities $b$, $c$ is defined as $ T_2$. The cavities lengths are defined as $L_a, L_b, L_c$. They should satisfy that the cavity coupling frequencies between each pair of adjacent cavities are identical. So the coupling frequency $\omega_c$ can be calculated as~\cite{PhysRevD.66.122004}
\begin{equation}
\omega_c=\frac{c\sqrt{T_1}}{2\sqrt{L_a L_b}}=\frac{c\sqrt{T_2}}{2\sqrt{L_b L_c}}\,.
\end{equation}
In the interaction picture, the optical part of the hamiltonian of the three coupled cavities can be written as
\begin{equation}
\mathcal{H}_{\rm opt}=\hbar\begin{bmatrix}
a^{\dagger}&b^{\dagger}&c^{\dagger}
\end{bmatrix}
\begin{bmatrix}
\omega_E &\omega_c & 0\\
\omega_c &\omega_E & \omega_c \\
0 &\omega_c &\omega_E
\end{bmatrix}
\begin{bmatrix}
a\\ b \\ c
\end{bmatrix}\,,
\end{equation}
The decoupled eigen-modes of the three coupled cavities, $n_0$, $n_{\pm}$, can be derived by diagonalising the matrix above, which gives
\begin{equation}
\begin{bmatrix}
a \\ b \\ c
\end{bmatrix}=
\begin{bmatrix}
\frac{1}{\sqrt{2}} &\frac{1}{2} &\frac{1}{2}\\
0 &-\frac{1}{\sqrt{2}} &\frac{1}{\sqrt{2}}\\
-\frac{1}{\sqrt{2}} & \frac{1}{2} &\frac{1}{2}
\end{bmatrix}
\begin{bmatrix}
n_{0} \\ n_{-} \\ n_{+}
\end{bmatrix}\,.
\end{equation} 
The three eigen-modes have eigen-frequencies, $\omega_E$ and $\omega_E\pm\sqrt{2}\omega_c$. For our purpose, we need
\begin{equation}
\omega_c=\sqrt{2}\omega_m\,.
\end{equation}
The hamiltonian describing the interactions in the system can be written as~\cite{walls2007quantum}
\begin{equation}\label{eq:Hamiton}
\begin{split}
\mathcal{H}_{\rm I}=&-\frac{i\hbar}{2}\Big[ \chi_0^a\frac{{n_0}^2+n_+n_-}{2}+{\chi_-^a}\frac{{n_-}^2}{4}+\chi_+^a\frac{{n_+}^2}{4}-\\
&\sqrt{2}\chi_0^b n_+ n_- +\chi_-^b\frac{{n_-}^2}{2}+\chi_+^b\frac{{n_+}^2}{2}\Big]+h.c.\,,
\end{split}
\end{equation}
where each constant, $\chi_0^a, \chi_-^a, \chi_+^a, \chi_0^b ,\chi_-^b, \chi_+^b$ is proportional to the second-order nonlinear susceptibility of crystal in cavity $a$ and $b$ and the amplitude of the pumps. 
By designing the power of pumps and crystal features satisfying  $\chi_0^a=2\sqrt{2}\chi_0^b$, the correlations between mode $n_+$ and $n_-$ can be coherently cancelled.
Thus Eq.~\eqref{eq:Hamiton} can be rewritten as
\begin{equation}\label{eq:Hamiton2}
\begin{split}
\mathcal{H}_{\rm I}=-\frac{i\hbar}{2} \left( g_0 {n_0^{\dagger}}^2+g_- {n_-^{\dagger}}^2 + g_+{n_{+}^{\dagger}}^2 \right)+h.c.
\end{split}
\end{equation}
where 
\begin{equation}
g_0=\frac{\chi_0^a}{2}\,,g_-=\frac{\chi_-^a+2\chi_-^b}{4}\,,g_+=\frac{\chi_+^a+2\chi_+^b}{4}\,.
\end{equation}
It is then straightforward to see the three pairs of independent interactions, which will give three modes of single mode squeezing.

\bibliography{bibliography}

\begin{thebibliography}{43}%
\makeatletter
\providecommand \@ifxundefined [1]{%
 \@ifx{#1\undefined}
}%
\providecommand \@ifnum [1]{%
 \ifnum #1\expandafter \@firstoftwo
 \else \expandafter \@secondoftwo
 \fi
}%
\providecommand \@ifx [1]{%
 \ifx #1\expandafter \@firstoftwo
 \else \expandafter \@secondoftwo
 \fi
}%
\providecommand \natexlab [1]{#1}%
\providecommand \enquote  [1]{``#1''}%
\providecommand \bibnamefont  [1]{#1}%
\providecommand \bibfnamefont [1]{#1}%
\providecommand \citenamefont [1]{#1}%
\providecommand \href@noop [0]{\@secondoftwo}%
\providecommand \href [0]{\begingroup \@sanitize@url \@href}%
\providecommand \@href[1]{\@@startlink{#1}\@@href}%
\providecommand \@@href[1]{\endgroup#1\@@endlink}%
\providecommand \@sanitize@url [0]{\catcode `\\12\catcode `\$12\catcode
  `\&12\catcode `\#12\catcode `\^12\catcode `\_12\catcode `\%12\relax}%
\providecommand \@@startlink[1]{}%
\providecommand \@@endlink[0]{}%
\providecommand \url  [0]{\begingroup\@sanitize@url \@url }%
\providecommand \@url [1]{\endgroup\@href {#1}{\urlprefix }}%
\providecommand \urlprefix  [0]{URL }%
\providecommand \Eprint [0]{\href }%
\providecommand \doibase [0]{http://dx.doi.org/}%
\providecommand \selectlanguage [0]{\@gobble}%
\providecommand \bibinfo  [0]{\@secondoftwo}%
\providecommand \bibfield  [0]{\@secondoftwo}%
\providecommand \translation [1]{[#1]}%
\providecommand \BibitemOpen [0]{}%
\providecommand \bibitemStop [0]{}%
\providecommand \bibitemNoStop [0]{.\EOS\space}%
\providecommand \EOS [0]{\spacefactor3000\relax}%
\providecommand \BibitemShut  [1]{\csname bibitem#1\endcsname}%
\let\auto@bib@innerbib\@empty
\bibitem [{\citenamefont {Drever}\ \emph {et~al.}(1983)\citenamefont {Drever},
  \citenamefont {Hall}, \citenamefont {Kowalski}, \citenamefont {Hough},
  \citenamefont {Ford}, \citenamefont {Munley},\ and\ \citenamefont
  {Ward}}]{DreverPDH}%
  \BibitemOpen
  \bibfield  {author} {\bibinfo {author} {\bibfnamefont {R.~W.~P.}\
  \bibnamefont {Drever}}, \bibinfo {author} {\bibfnamefont {J.~L.}\
  \bibnamefont {Hall}}, \bibinfo {author} {\bibfnamefont {F.~V.}\ \bibnamefont
  {Kowalski}}, \bibinfo {author} {\bibfnamefont {J.}~\bibnamefont {Hough}},
  \bibinfo {author} {\bibfnamefont {G.~M.}\ \bibnamefont {Ford}}, \bibinfo
  {author} {\bibfnamefont {A.~J.}\ \bibnamefont {Munley}}, \ and\ \bibinfo
  {author} {\bibfnamefont {H.}~\bibnamefont {Ward}},\ }\href {\doibase
  10.1007/BF00702605} {\bibfield  {journal} {\bibinfo  {journal} {Applied
  Physics B}\ }\textbf {\bibinfo {volume} {31}},\ \bibinfo {pages} {97}
  (\bibinfo {year} {1983})}\BibitemShut {NoStop}%
\bibitem [{\citenamefont {Black}(2001)}]{Black_PDH_2001}%
  \BibitemOpen
  \bibfield  {author} {\bibinfo {author} {\bibfnamefont {E.~D.}\ \bibnamefont
  {Black}},\ }\href {\doibase 10.1119/1.1286663} {\bibfield  {journal}
  {\bibinfo  {journal} {American Journal of Physics}\ }\textbf {\bibinfo
  {volume} {69}},\ \bibinfo {pages} {79} (\bibinfo {year} {2001})}\BibitemShut
  {NoStop}%
\bibitem [{\citenamefont {Takahashi}\ and\ \citenamefont {the
  TAMA~Collaboration}(2004)}]{Takahashi_2004}%
  \BibitemOpen
  \bibfield  {author} {\bibinfo {author} {\bibfnamefont {R.}~\bibnamefont
  {Takahashi}}\ and\ \bibinfo {author} {\bibnamefont {the
  TAMA~Collaboration}},\ }\href {\doibase 10.1088/0264-9381/21/5/004}
  {\bibfield  {journal} {\bibinfo  {journal} {Classical and Quantum Gravity}\
  }\textbf {\bibinfo {volume} {21}},\ \bibinfo {pages} {S403} (\bibinfo {year}
  {2004})}\BibitemShut {NoStop}%
\bibitem [{\citenamefont {(for~the LIGO
  Scientific~Collaboration)}(2008)}]{Sigg_2008}%
  \BibitemOpen
  \bibfield  {author} {\bibinfo {author} {\bibfnamefont {D.~S.}\ \bibnamefont
  {(for~the LIGO Scientific~Collaboration)}},\ }\href {\doibase
  10.1088/0264-9381/25/11/114041} {\bibfield  {journal} {\bibinfo  {journal}
  {Classical and Quantum Gravity}\ }\textbf {\bibinfo {volume} {25}},\ \bibinfo
  {pages} {114041} (\bibinfo {year} {2008})}\BibitemShut {NoStop}%
\bibitem [{\citenamefont {Aasi}\ \emph
  {et~al.}(2015{\natexlab{a}})\citenamefont {Aasi}, \citenamefont {Abbott},
  \citenamefont {Abbott}, \citenamefont {Abbott}, \citenamefont {Abernathy},
  \citenamefont {Ackley}, \citenamefont {Adams}, \citenamefont {Adams},
  \citenamefont {Addesso}, \citenamefont {Adhikari},\ and\ \citenamefont
  {Collaboration}}]{LIGO2015}%
  \BibitemOpen
  \bibfield  {author} {\bibinfo {author} {\bibfnamefont {J.}~\bibnamefont
  {Aasi}}, \bibinfo {author} {\bibfnamefont {B.~P.}\ \bibnamefont {Abbott}},
  \bibinfo {author} {\bibfnamefont {R.}~\bibnamefont {Abbott}}, \bibinfo
  {author} {\bibfnamefont {T.}~\bibnamefont {Abbott}}, \bibinfo {author}
  {\bibfnamefont {M.~R.}\ \bibnamefont {Abernathy}}, \bibinfo {author}
  {\bibfnamefont {K.}~\bibnamefont {Ackley}}, \bibinfo {author} {\bibfnamefont
  {C.}~\bibnamefont {Adams}}, \bibinfo {author} {\bibfnamefont
  {T.}~\bibnamefont {Adams}}, \bibinfo {author} {\bibfnamefont
  {P.}~\bibnamefont {Addesso}}, \bibinfo {author} {\bibfnamefont {R.~X.}\
  \bibnamefont {Adhikari}}, \ and\ \bibinfo {author} {\bibfnamefont {L.~S.}\
  \bibnamefont {Collaboration}},\ }\href {\doibase
  10.1088/0264-9381/32/7/074001} {\bibfield  {journal} {\bibinfo  {journal}
  {Classical and Quantum Gravity}\ }\textbf {\bibinfo {volume} {32}},\ \bibinfo
  {pages} {074001} (\bibinfo {year} {2015}{\natexlab{a}})}\BibitemShut
  {NoStop}%
\bibitem [{\citenamefont {Acernese}\ \emph {et~al.}(2006)\citenamefont
  {Acernese}, \citenamefont {Amico}, \citenamefont {Al-Shourbagy},
  \citenamefont {Aoudia}, \citenamefont {Avino}, \citenamefont {Babusci},
  \citenamefont {Ballardin}, \citenamefont {Barone}, \citenamefont {Barsotti},
  \citenamefont {Barsuglia}, \citenamefont {Beauville}, \citenamefont
  {Bizouard}, \citenamefont {Boccara}, \citenamefont {Bondu}, \citenamefont
  {Bosi}, \citenamefont {Bradaschia}, \citenamefont {Birindelli}, \citenamefont
  {Braccini}, \citenamefont {Brillet}, \citenamefont {Brisson}, \citenamefont
  {Brocco}, \citenamefont {Buskulic}, \citenamefont {Calloni}, \citenamefont
  {Campagna}, \citenamefont {Cavalier}, \citenamefont {Cavalieri},
  \citenamefont {Cella}, \citenamefont {Chassande-Mottin}, \citenamefont
  {Corda}, \citenamefont {Clapson}, \citenamefont {Cleva}, \citenamefont
  {Coulon}, \citenamefont {Cuoco}, \citenamefont {Dattilo}, \citenamefont
  {Davier}, \citenamefont {Rosa}, \citenamefont {Fiore}, \citenamefont
  {Virgilio}, \citenamefont {Dujardin}, \citenamefont {Eleuteri}, \citenamefont
  {Enard}, \citenamefont {Ferrante}, \citenamefont {Fidecaro}, \citenamefont
  {Fiori}, \citenamefont {Flaminio}, \citenamefont {Fournier}, \citenamefont
  {Francois}, \citenamefont {Frasca}, \citenamefont {Frasconi}, \citenamefont
  {Freise}, \citenamefont {Gammaitoni}, \citenamefont {Gennai}, \citenamefont
  {Giazotto}, \citenamefont {Giordano}, \citenamefont {Giordano}, \citenamefont
  {Gouaty}, \citenamefont {Grosjean}, \citenamefont {Guidi}, \citenamefont
  {Hebri}, \citenamefont {Heitmann}, \citenamefont {Hello}, \citenamefont
  {Holloway}, \citenamefont {Karkar}, \citenamefont {Kreckelbergh},
  \citenamefont {Penna}, \citenamefont {Letendre}, \citenamefont {Lorenzini},
  \citenamefont {Loriette}, \citenamefont {Loupias}, \citenamefont {Losurdo},
  \citenamefont {Mackowski}, \citenamefont {Majorana}, \citenamefont {Man},
  \citenamefont {Mantovani}, \citenamefont {Marchesoni}, \citenamefont
  {Marion}, \citenamefont {Marque}, \citenamefont {Martelli}, \citenamefont
  {Masserot}, \citenamefont {Mazzoni}, \citenamefont {Milano}, \citenamefont
  {Moins}, \citenamefont {Moreau}, \citenamefont {Morgado}, \citenamefont
  {Mours}, \citenamefont {Pai}, \citenamefont {Palomba}, \citenamefont
  {Paoletti}, \citenamefont {Pardi}, \citenamefont {Pasqualetti}, \citenamefont
  {Passaquieti}, \citenamefont {Passuello}, \citenamefont {Perniola},
  \citenamefont {Piergiovanni}, \citenamefont {Pinard}, \citenamefont
  {Poggiani}, \citenamefont {Punturo}, \citenamefont {Puppo}, \citenamefont
  {Qipiani}, \citenamefont {Rapagnani}, \citenamefont {Reita}, \citenamefont
  {Remillieux}, \citenamefont {Ricci}, \citenamefont {Ricciardi}, \citenamefont
  {Ruggi}, \citenamefont {Russo}, \citenamefont {Solimeno}, \citenamefont
  {Spallicci}, \citenamefont {Stanga}, \citenamefont {Taddei}, \citenamefont
  {Tonelli}, \citenamefont {Toncelli}, \citenamefont {Tournefier},
  \citenamefont {Travasso}, \citenamefont {Vajente}, \citenamefont {Verkindt},
  \citenamefont {Vetrano}, \citenamefont {Vicer{\'{e}}}, \citenamefont {Vinet},
  \citenamefont {Vocca}, \citenamefont {Yvert},\ and\ \citenamefont
  {Zhang}}]{Acernese_2006}%
  \BibitemOpen
  \bibfield  {author} {\bibinfo {author} {\bibfnamefont {F.}~\bibnamefont
  {Acernese}}, \bibinfo {author} {\bibfnamefont {P.}~\bibnamefont {Amico}},
  \bibinfo {author} {\bibfnamefont {M.}~\bibnamefont {Al-Shourbagy}}, \bibinfo
  {author} {\bibfnamefont {S.}~\bibnamefont {Aoudia}}, \bibinfo {author}
  {\bibfnamefont {S.}~\bibnamefont {Avino}}, \bibinfo {author} {\bibfnamefont
  {D.}~\bibnamefont {Babusci}}, \bibinfo {author} {\bibfnamefont
  {G.}~\bibnamefont {Ballardin}}, \bibinfo {author} {\bibfnamefont
  {F.}~\bibnamefont {Barone}}, \bibinfo {author} {\bibfnamefont
  {L.}~\bibnamefont {Barsotti}}, \bibinfo {author} {\bibfnamefont
  {M.}~\bibnamefont {Barsuglia}}, \bibinfo {author} {\bibfnamefont
  {F.}~\bibnamefont {Beauville}}, \bibinfo {author} {\bibfnamefont {M.~A.}\
  \bibnamefont {Bizouard}}, \bibinfo {author} {\bibfnamefont {C.}~\bibnamefont
  {Boccara}}, \bibinfo {author} {\bibfnamefont {F.}~\bibnamefont {Bondu}},
  \bibinfo {author} {\bibfnamefont {L.}~\bibnamefont {Bosi}}, \bibinfo {author}
  {\bibfnamefont {C.}~\bibnamefont {Bradaschia}}, \bibinfo {author}
  {\bibfnamefont {S.}~\bibnamefont {Birindelli}}, \bibinfo {author}
  {\bibfnamefont {S.}~\bibnamefont {Braccini}}, \bibinfo {author}
  {\bibfnamefont {A.}~\bibnamefont {Brillet}}, \bibinfo {author} {\bibfnamefont
  {V.}~\bibnamefont {Brisson}}, \bibinfo {author} {\bibfnamefont
  {L.}~\bibnamefont {Brocco}}, \bibinfo {author} {\bibfnamefont
  {D.}~\bibnamefont {Buskulic}}, \bibinfo {author} {\bibfnamefont
  {E.}~\bibnamefont {Calloni}}, \bibinfo {author} {\bibfnamefont
  {E.}~\bibnamefont {Campagna}}, \bibinfo {author} {\bibfnamefont
  {F.}~\bibnamefont {Cavalier}}, \bibinfo {author} {\bibfnamefont
  {R.}~\bibnamefont {Cavalieri}}, \bibinfo {author} {\bibfnamefont
  {G.}~\bibnamefont {Cella}}, \bibinfo {author} {\bibfnamefont
  {E.}~\bibnamefont {Chassande-Mottin}}, \bibinfo {author} {\bibfnamefont
  {C.}~\bibnamefont {Corda}}, \bibinfo {author} {\bibfnamefont {A.-C.}\
  \bibnamefont {Clapson}}, \bibinfo {author} {\bibfnamefont {F.}~\bibnamefont
  {Cleva}}, \bibinfo {author} {\bibfnamefont {J.-P.}\ \bibnamefont {Coulon}},
  \bibinfo {author} {\bibfnamefont {E.}~\bibnamefont {Cuoco}}, \bibinfo
  {author} {\bibfnamefont {V.}~\bibnamefont {Dattilo}}, \bibinfo {author}
  {\bibfnamefont {M.}~\bibnamefont {Davier}}, \bibinfo {author} {\bibfnamefont
  {R.~D.}\ \bibnamefont {Rosa}}, \bibinfo {author} {\bibfnamefont {L.~D.}\
  \bibnamefont {Fiore}}, \bibinfo {author} {\bibfnamefont {A.~D.}\ \bibnamefont
  {Virgilio}}, \bibinfo {author} {\bibfnamefont {B.}~\bibnamefont {Dujardin}},
  \bibinfo {author} {\bibfnamefont {A.}~\bibnamefont {Eleuteri}}, \bibinfo
  {author} {\bibfnamefont {D.}~\bibnamefont {Enard}}, \bibinfo {author}
  {\bibfnamefont {I.}~\bibnamefont {Ferrante}}, \bibinfo {author}
  {\bibfnamefont {F.}~\bibnamefont {Fidecaro}}, \bibinfo {author}
  {\bibfnamefont {I.}~\bibnamefont {Fiori}}, \bibinfo {author} {\bibfnamefont
  {R.}~\bibnamefont {Flaminio}}, \bibinfo {author} {\bibfnamefont {J.-D.}\
  \bibnamefont {Fournier}}, \bibinfo {author} {\bibfnamefont {O.}~\bibnamefont
  {Francois}}, \bibinfo {author} {\bibfnamefont {S.}~\bibnamefont {Frasca}},
  \bibinfo {author} {\bibfnamefont {F.}~\bibnamefont {Frasconi}}, \bibinfo
  {author} {\bibfnamefont {A.}~\bibnamefont {Freise}}, \bibinfo {author}
  {\bibfnamefont {L.}~\bibnamefont {Gammaitoni}}, \bibinfo {author}
  {\bibfnamefont {A.}~\bibnamefont {Gennai}}, \bibinfo {author} {\bibfnamefont
  {A.}~\bibnamefont {Giazotto}}, \bibinfo {author} {\bibfnamefont
  {G.}~\bibnamefont {Giordano}}, \bibinfo {author} {\bibfnamefont
  {L.}~\bibnamefont {Giordano}}, \bibinfo {author} {\bibfnamefont
  {R.}~\bibnamefont {Gouaty}}, \bibinfo {author} {\bibfnamefont
  {D.}~\bibnamefont {Grosjean}}, \bibinfo {author} {\bibfnamefont
  {G.}~\bibnamefont {Guidi}}, \bibinfo {author} {\bibfnamefont
  {S.}~\bibnamefont {Hebri}}, \bibinfo {author} {\bibfnamefont
  {H.}~\bibnamefont {Heitmann}}, \bibinfo {author} {\bibfnamefont
  {P.}~\bibnamefont {Hello}}, \bibinfo {author} {\bibfnamefont
  {L.}~\bibnamefont {Holloway}}, \bibinfo {author} {\bibfnamefont
  {S.}~\bibnamefont {Karkar}}, \bibinfo {author} {\bibfnamefont
  {S.}~\bibnamefont {Kreckelbergh}}, \bibinfo {author} {\bibfnamefont {P.~L.}\
  \bibnamefont {Penna}}, \bibinfo {author} {\bibfnamefont {N.}~\bibnamefont
  {Letendre}}, \bibinfo {author} {\bibfnamefont {M.}~\bibnamefont {Lorenzini}},
  \bibinfo {author} {\bibfnamefont {V.}~\bibnamefont {Loriette}}, \bibinfo
  {author} {\bibfnamefont {M.}~\bibnamefont {Loupias}}, \bibinfo {author}
  {\bibfnamefont {G.}~\bibnamefont {Losurdo}}, \bibinfo {author} {\bibfnamefont
  {J.~M.}\ \bibnamefont {Mackowski}}, \bibinfo {author} {\bibfnamefont
  {E.}~\bibnamefont {Majorana}}, \bibinfo {author} {\bibfnamefont {C.~N.}\
  \bibnamefont {Man}}, \bibinfo {author} {\bibfnamefont {M.}~\bibnamefont
  {Mantovani}}, \bibinfo {author} {\bibfnamefont {F.}~\bibnamefont
  {Marchesoni}}, \bibinfo {author} {\bibfnamefont {F.}~\bibnamefont {Marion}},
  \bibinfo {author} {\bibfnamefont {J.}~\bibnamefont {Marque}}, \bibinfo
  {author} {\bibfnamefont {F.}~\bibnamefont {Martelli}}, \bibinfo {author}
  {\bibfnamefont {A.}~\bibnamefont {Masserot}}, \bibinfo {author}
  {\bibfnamefont {M.}~\bibnamefont {Mazzoni}}, \bibinfo {author} {\bibfnamefont
  {L.}~\bibnamefont {Milano}}, \bibinfo {author} {\bibfnamefont
  {C.}~\bibnamefont {Moins}}, \bibinfo {author} {\bibfnamefont
  {J.}~\bibnamefont {Moreau}}, \bibinfo {author} {\bibfnamefont
  {N.}~\bibnamefont {Morgado}}, \bibinfo {author} {\bibfnamefont
  {B.}~\bibnamefont {Mours}}, \bibinfo {author} {\bibfnamefont
  {A.}~\bibnamefont {Pai}}, \bibinfo {author} {\bibfnamefont {C.}~\bibnamefont
  {Palomba}}, \bibinfo {author} {\bibfnamefont {F.}~\bibnamefont {Paoletti}},
  \bibinfo {author} {\bibfnamefont {S.}~\bibnamefont {Pardi}}, \bibinfo
  {author} {\bibfnamefont {A.}~\bibnamefont {Pasqualetti}}, \bibinfo {author}
  {\bibfnamefont {R.}~\bibnamefont {Passaquieti}}, \bibinfo {author}
  {\bibfnamefont {D.}~\bibnamefont {Passuello}}, \bibinfo {author}
  {\bibfnamefont {B.}~\bibnamefont {Perniola}}, \bibinfo {author}
  {\bibfnamefont {F.}~\bibnamefont {Piergiovanni}}, \bibinfo {author}
  {\bibfnamefont {L.}~\bibnamefont {Pinard}}, \bibinfo {author} {\bibfnamefont
  {R.}~\bibnamefont {Poggiani}}, \bibinfo {author} {\bibfnamefont
  {M.}~\bibnamefont {Punturo}}, \bibinfo {author} {\bibfnamefont
  {P.}~\bibnamefont {Puppo}}, \bibinfo {author} {\bibfnamefont
  {K.}~\bibnamefont {Qipiani}}, \bibinfo {author} {\bibfnamefont
  {P.}~\bibnamefont {Rapagnani}}, \bibinfo {author} {\bibfnamefont
  {V.}~\bibnamefont {Reita}}, \bibinfo {author} {\bibfnamefont
  {A.}~\bibnamefont {Remillieux}}, \bibinfo {author} {\bibfnamefont
  {F.}~\bibnamefont {Ricci}}, \bibinfo {author} {\bibfnamefont
  {I.}~\bibnamefont {Ricciardi}}, \bibinfo {author} {\bibfnamefont
  {P.}~\bibnamefont {Ruggi}}, \bibinfo {author} {\bibfnamefont
  {G.}~\bibnamefont {Russo}}, \bibinfo {author} {\bibfnamefont
  {S.}~\bibnamefont {Solimeno}}, \bibinfo {author} {\bibfnamefont
  {A.}~\bibnamefont {Spallicci}}, \bibinfo {author} {\bibfnamefont
  {R.}~\bibnamefont {Stanga}}, \bibinfo {author} {\bibfnamefont
  {R.}~\bibnamefont {Taddei}}, \bibinfo {author} {\bibfnamefont
  {M.}~\bibnamefont {Tonelli}}, \bibinfo {author} {\bibfnamefont
  {A.}~\bibnamefont {Toncelli}}, \bibinfo {author} {\bibfnamefont
  {E.}~\bibnamefont {Tournefier}}, \bibinfo {author} {\bibfnamefont
  {F.}~\bibnamefont {Travasso}}, \bibinfo {author} {\bibfnamefont
  {G.}~\bibnamefont {Vajente}}, \bibinfo {author} {\bibfnamefont
  {D.}~\bibnamefont {Verkindt}}, \bibinfo {author} {\bibfnamefont
  {F.}~\bibnamefont {Vetrano}}, \bibinfo {author} {\bibfnamefont
  {A.}~\bibnamefont {Vicer{\'{e}}}}, \bibinfo {author} {\bibfnamefont {J.-Y.}\
  \bibnamefont {Vinet}}, \bibinfo {author} {\bibfnamefont {H.}~\bibnamefont
  {Vocca}}, \bibinfo {author} {\bibfnamefont {M.}~\bibnamefont {Yvert}}, \ and\
  \bibinfo {author} {\bibfnamefont {Z.}~\bibnamefont {Zhang}},\ }\href
  {\doibase 10.1088/0264-9381/23/8/s09} {\bibfield  {journal} {\bibinfo
  {journal} {Classical and Quantum Gravity}\ }\textbf {\bibinfo {volume}
  {23}},\ \bibinfo {pages} {S63} (\bibinfo {year} {2006})}\BibitemShut
  {NoStop}%
\bibitem [{\citenamefont {Acernese}\ \emph {et~al.}(2014)\citenamefont
  {Acernese}, \citenamefont {Agathos}, \citenamefont {Agatsuma}, \citenamefont
  {Aisa}, \citenamefont {Allemandou}, \citenamefont {Allocca}, \citenamefont
  {Amarni}, \citenamefont {Astone}, \citenamefont {Balestri}, \citenamefont
  {Ballardin}, \citenamefont {Barone}, \citenamefont {Baronick}, \citenamefont
  {Barsuglia}, \citenamefont {Basti}, \citenamefont {Basti}, \citenamefont
  {Bauer}, \citenamefont {Bavigadda}, \citenamefont {Bejger}, \citenamefont
  {Beker}, \citenamefont {Belczynski}, \citenamefont {Bersanetti},
  \citenamefont {Bertolini}, \citenamefont {Bitossi}, \citenamefont {Bizouard},
  \citenamefont {Bloemen}, \citenamefont {Blom}, \citenamefont {Boer},
  \citenamefont {Bogaert}, \citenamefont {Bondi}, \citenamefont {Bondu},
  \citenamefont {Bonelli}, \citenamefont {Bonnand}, \citenamefont {Boschi},
  \citenamefont {Bosi}, \citenamefont {Bouedo}, \citenamefont {Bradaschia},
  \citenamefont {Branchesi}, \citenamefont {Briant}, \citenamefont {Brillet},
  \citenamefont {Brisson}, \citenamefont {Bulik}, \citenamefont {Bulten},
  \citenamefont {Buskulic}, \citenamefont {Buy}, \citenamefont {Cagnoli},
  \citenamefont {Calloni}, \citenamefont {Campeggi}, \citenamefont {Canuel},
  \citenamefont {Carbognani}, \citenamefont {Cavalier}, \citenamefont
  {Cavalieri}, \citenamefont {Cella}, \citenamefont {Cesarini}, \citenamefont
  {Chassande-Mottin}, \citenamefont {Chincarini}, \citenamefont {Chiummo},
  \citenamefont {Chua}, \citenamefont {Cleva}, \citenamefont {Coccia},
  \citenamefont {Cohadon}, \citenamefont {Colla}, \citenamefont {Colombini},
  \citenamefont {Conte}, \citenamefont {Coulon}, \citenamefont {Cuoco},
  \citenamefont {Dalmaz}, \citenamefont {D'Antonio}, \citenamefont {Dattilo},
  \citenamefont {Davier}, \citenamefont {Day}, \citenamefont {Debreczeni},
  \citenamefont {Degallaix}, \citenamefont {Del{\'{e}}glise}, \citenamefont
  {Pozzo}, \citenamefont {Dereli}, \citenamefont {Rosa}, \citenamefont {Fiore},
  \citenamefont {Lieto}, \citenamefont {Virgilio}, \citenamefont {Doets},
  \citenamefont {Dolique}, \citenamefont {Drago}, \citenamefont {Ducrot},
  \citenamefont {Endr{\H{o}}czi}, \citenamefont {Fafone}, \citenamefont
  {Farinon}, \citenamefont {Ferrante}, \citenamefont {Ferrini}, \citenamefont
  {Fidecaro}, \citenamefont {Fiori}, \citenamefont {Flaminio}, \citenamefont
  {Fournier}, \citenamefont {Franco}, \citenamefont {Frasca}, \citenamefont
  {Frasconi}, \citenamefont {Gammaitoni}, \citenamefont {Garufi}, \citenamefont
  {Gaspard}, \citenamefont {Gatto}, \citenamefont {Gemme}, \citenamefont
  {Gendre}, \citenamefont {Genin}, \citenamefont {Gennai}, \citenamefont
  {Ghosh}, \citenamefont {Giacobone}, \citenamefont {Giazotto}, \citenamefont
  {Gouaty}, \citenamefont {Granata}, \citenamefont {Greco}, \citenamefont
  {Groot}, \citenamefont {Guidi}, \citenamefont {Harms}, \citenamefont
  {Heidmann}, \citenamefont {Heitmann}, \citenamefont {Hello}, \citenamefont
  {Hemming}, \citenamefont {Hennes}, \citenamefont {Hofman}, \citenamefont
  {Jaranowski}, \citenamefont {Jonker}, \citenamefont {Kasprzack},
  \citenamefont {K{\'{e}}f{\'{e}}lian}, \citenamefont {Kowalska}, \citenamefont
  {Kraan}, \citenamefont {Kr{\'{o}}lak}, \citenamefont {Kutynia}, \citenamefont
  {Lazzaro}, \citenamefont {Leonardi}, \citenamefont {Leroy}, \citenamefont
  {Letendre}, \citenamefont {Li}, \citenamefont {Lieunard}, \citenamefont
  {Lorenzini}, \citenamefont {Loriette}, \citenamefont {Losurdo}, \citenamefont
  {Magazz{\`{u}}}, \citenamefont {Majorana}, \citenamefont {Maksimovic},
  \citenamefont {Malvezzi}, \citenamefont {Man}, \citenamefont {Mangano},
  \citenamefont {Mantovani}, \citenamefont {Marchesoni}, \citenamefont
  {Marion}, \citenamefont {Marque}, \citenamefont {Martelli}, \citenamefont
  {Martellini}, \citenamefont {Masserot}, \citenamefont {Meacher},
  \citenamefont {Meidam}, \citenamefont {Mezzani}, \citenamefont {Michel},
  \citenamefont {Milano}, \citenamefont {Minenkov}, \citenamefont {Moggi},
  \citenamefont {Mohan}, \citenamefont {Montani}, \citenamefont {Morgado},
  \citenamefont {Mours}, \citenamefont {Mul}, \citenamefont {Nagy},
  \citenamefont {Nardecchia}, \citenamefont {Naticchioni}, \citenamefont
  {Nelemans}, \citenamefont {Neri}, \citenamefont {Neri}, \citenamefont
  {Nocera}, \citenamefont {Pacaud}, \citenamefont {Palomba}, \citenamefont
  {Paoletti}, \citenamefont {Paoli}, \citenamefont {Pasqualetti}, \citenamefont
  {Passaquieti}, \citenamefont {Passuello}, \citenamefont {Perciballi},
  \citenamefont {Petit}, \citenamefont {Pichot}, \citenamefont {Piergiovanni},
  \citenamefont {Pillant}, \citenamefont {Piluso}, \citenamefont {Pinard},
  \citenamefont {Poggiani}, \citenamefont {Prijatelj}, \citenamefont {Prodi},
  \citenamefont {Punturo}, \citenamefont {Puppo}, \citenamefont {Rabeling},
  \citenamefont {R{\'{a}}cz}, \citenamefont {Rapagnani}, \citenamefont
  {Razzano}, \citenamefont {Re}, \citenamefont {Regimbau}, \citenamefont
  {Ricci}, \citenamefont {Robinet}, \citenamefont {Rocchi}, \citenamefont
  {Rolland}, \citenamefont {Romano}, \citenamefont {Rosi{\'{n}}ska},
  \citenamefont {Ruggi}, \citenamefont {Saracco}, \citenamefont {Sassolas},
  \citenamefont {Schimmel}, \citenamefont {Sentenac}, \citenamefont {Sequino},
  \citenamefont {Shah}, \citenamefont {Siellez}, \citenamefont {Straniero},
  \citenamefont {Swinkels}, \citenamefont {Tacca}, \citenamefont {Tonelli},
  \citenamefont {Travasso}, \citenamefont {Turconi}, \citenamefont {Vajente},
  \citenamefont {van Bakel}, \citenamefont {van Beuzekom}, \citenamefont
  {van~den Brand}, \citenamefont {Broeck}, \citenamefont {van~der Sluys},
  \citenamefont {van Heijningen}, \citenamefont {Vas{\'{u}}th}, \citenamefont
  {Vedovato}, \citenamefont {Veitch}, \citenamefont {Verkindt}, \citenamefont
  {Vetrano}, \citenamefont {Vicer{\'{e}}}, \citenamefont {Vinet}, \citenamefont
  {Visser}, \citenamefont {Vocca}, \citenamefont {Ward}, \citenamefont {Was},
  \citenamefont {Wei}, \citenamefont {Yvert}, \citenamefont {{\.{z}}ny},\ and\
  \citenamefont {Zendri}}]{Acernese_2014}%
  \BibitemOpen
  \bibfield  {author} {\bibinfo {author} {\bibfnamefont {F.}~\bibnamefont
  {Acernese}}, \bibinfo {author} {\bibfnamefont {M.}~\bibnamefont {Agathos}},
  \bibinfo {author} {\bibfnamefont {K.}~\bibnamefont {Agatsuma}}, \bibinfo
  {author} {\bibfnamefont {D.}~\bibnamefont {Aisa}}, \bibinfo {author}
  {\bibfnamefont {N.}~\bibnamefont {Allemandou}}, \bibinfo {author}
  {\bibfnamefont {A.}~\bibnamefont {Allocca}}, \bibinfo {author} {\bibfnamefont
  {J.}~\bibnamefont {Amarni}}, \bibinfo {author} {\bibfnamefont
  {P.}~\bibnamefont {Astone}}, \bibinfo {author} {\bibfnamefont
  {G.}~\bibnamefont {Balestri}}, \bibinfo {author} {\bibfnamefont
  {G.}~\bibnamefont {Ballardin}}, \bibinfo {author} {\bibfnamefont
  {F.}~\bibnamefont {Barone}}, \bibinfo {author} {\bibfnamefont {J.-P.}\
  \bibnamefont {Baronick}}, \bibinfo {author} {\bibfnamefont {M.}~\bibnamefont
  {Barsuglia}}, \bibinfo {author} {\bibfnamefont {A.}~\bibnamefont {Basti}},
  \bibinfo {author} {\bibfnamefont {F.}~\bibnamefont {Basti}}, \bibinfo
  {author} {\bibfnamefont {T.~S.}\ \bibnamefont {Bauer}}, \bibinfo {author}
  {\bibfnamefont {V.}~\bibnamefont {Bavigadda}}, \bibinfo {author}
  {\bibfnamefont {M.}~\bibnamefont {Bejger}}, \bibinfo {author} {\bibfnamefont
  {M.~G.}\ \bibnamefont {Beker}}, \bibinfo {author} {\bibfnamefont
  {C.}~\bibnamefont {Belczynski}}, \bibinfo {author} {\bibfnamefont
  {D.}~\bibnamefont {Bersanetti}}, \bibinfo {author} {\bibfnamefont
  {A.}~\bibnamefont {Bertolini}}, \bibinfo {author} {\bibfnamefont
  {M.}~\bibnamefont {Bitossi}}, \bibinfo {author} {\bibfnamefont {M.~A.}\
  \bibnamefont {Bizouard}}, \bibinfo {author} {\bibfnamefont {S.}~\bibnamefont
  {Bloemen}}, \bibinfo {author} {\bibfnamefont {M.}~\bibnamefont {Blom}},
  \bibinfo {author} {\bibfnamefont {M.}~\bibnamefont {Boer}}, \bibinfo {author}
  {\bibfnamefont {G.}~\bibnamefont {Bogaert}}, \bibinfo {author} {\bibfnamefont
  {D.}~\bibnamefont {Bondi}}, \bibinfo {author} {\bibfnamefont
  {F.}~\bibnamefont {Bondu}}, \bibinfo {author} {\bibfnamefont
  {L.}~\bibnamefont {Bonelli}}, \bibinfo {author} {\bibfnamefont
  {R.}~\bibnamefont {Bonnand}}, \bibinfo {author} {\bibfnamefont
  {V.}~\bibnamefont {Boschi}}, \bibinfo {author} {\bibfnamefont
  {L.}~\bibnamefont {Bosi}}, \bibinfo {author} {\bibfnamefont {T.}~\bibnamefont
  {Bouedo}}, \bibinfo {author} {\bibfnamefont {C.}~\bibnamefont {Bradaschia}},
  \bibinfo {author} {\bibfnamefont {M.}~\bibnamefont {Branchesi}}, \bibinfo
  {author} {\bibfnamefont {T.}~\bibnamefont {Briant}}, \bibinfo {author}
  {\bibfnamefont {A.}~\bibnamefont {Brillet}}, \bibinfo {author} {\bibfnamefont
  {V.}~\bibnamefont {Brisson}}, \bibinfo {author} {\bibfnamefont
  {T.}~\bibnamefont {Bulik}}, \bibinfo {author} {\bibfnamefont {H.~J.}\
  \bibnamefont {Bulten}}, \bibinfo {author} {\bibfnamefont {D.}~\bibnamefont
  {Buskulic}}, \bibinfo {author} {\bibfnamefont {C.}~\bibnamefont {Buy}},
  \bibinfo {author} {\bibfnamefont {G.}~\bibnamefont {Cagnoli}}, \bibinfo
  {author} {\bibfnamefont {E.}~\bibnamefont {Calloni}}, \bibinfo {author}
  {\bibfnamefont {C.}~\bibnamefont {Campeggi}}, \bibinfo {author}
  {\bibfnamefont {B.}~\bibnamefont {Canuel}}, \bibinfo {author} {\bibfnamefont
  {F.}~\bibnamefont {Carbognani}}, \bibinfo {author} {\bibfnamefont
  {F.}~\bibnamefont {Cavalier}}, \bibinfo {author} {\bibfnamefont
  {R.}~\bibnamefont {Cavalieri}}, \bibinfo {author} {\bibfnamefont
  {G.}~\bibnamefont {Cella}}, \bibinfo {author} {\bibfnamefont
  {E.}~\bibnamefont {Cesarini}}, \bibinfo {author} {\bibfnamefont
  {E.}~\bibnamefont {Chassande-Mottin}}, \bibinfo {author} {\bibfnamefont
  {A.}~\bibnamefont {Chincarini}}, \bibinfo {author} {\bibfnamefont
  {A.}~\bibnamefont {Chiummo}}, \bibinfo {author} {\bibfnamefont
  {S.}~\bibnamefont {Chua}}, \bibinfo {author} {\bibfnamefont {F.}~\bibnamefont
  {Cleva}}, \bibinfo {author} {\bibfnamefont {E.}~\bibnamefont {Coccia}},
  \bibinfo {author} {\bibfnamefont {P.-F.}\ \bibnamefont {Cohadon}}, \bibinfo
  {author} {\bibfnamefont {A.}~\bibnamefont {Colla}}, \bibinfo {author}
  {\bibfnamefont {M.}~\bibnamefont {Colombini}}, \bibinfo {author}
  {\bibfnamefont {A.}~\bibnamefont {Conte}}, \bibinfo {author} {\bibfnamefont
  {J.-P.}\ \bibnamefont {Coulon}}, \bibinfo {author} {\bibfnamefont
  {E.}~\bibnamefont {Cuoco}}, \bibinfo {author} {\bibfnamefont
  {A.}~\bibnamefont {Dalmaz}}, \bibinfo {author} {\bibfnamefont
  {S.}~\bibnamefont {D'Antonio}}, \bibinfo {author} {\bibfnamefont
  {V.}~\bibnamefont {Dattilo}}, \bibinfo {author} {\bibfnamefont
  {M.}~\bibnamefont {Davier}}, \bibinfo {author} {\bibfnamefont
  {R.}~\bibnamefont {Day}}, \bibinfo {author} {\bibfnamefont {G.}~\bibnamefont
  {Debreczeni}}, \bibinfo {author} {\bibfnamefont {J.}~\bibnamefont
  {Degallaix}}, \bibinfo {author} {\bibfnamefont {S.}~\bibnamefont
  {Del{\'{e}}glise}}, \bibinfo {author} {\bibfnamefont {W.~D.}\ \bibnamefont
  {Pozzo}}, \bibinfo {author} {\bibfnamefont {H.}~\bibnamefont {Dereli}},
  \bibinfo {author} {\bibfnamefont {R.~D.}\ \bibnamefont {Rosa}}, \bibinfo
  {author} {\bibfnamefont {L.~D.}\ \bibnamefont {Fiore}}, \bibinfo {author}
  {\bibfnamefont {A.~D.}\ \bibnamefont {Lieto}}, \bibinfo {author}
  {\bibfnamefont {A.~D.}\ \bibnamefont {Virgilio}}, \bibinfo {author}
  {\bibfnamefont {M.}~\bibnamefont {Doets}}, \bibinfo {author} {\bibfnamefont
  {V.}~\bibnamefont {Dolique}}, \bibinfo {author} {\bibfnamefont
  {M.}~\bibnamefont {Drago}}, \bibinfo {author} {\bibfnamefont
  {M.}~\bibnamefont {Ducrot}}, \bibinfo {author} {\bibfnamefont
  {G.}~\bibnamefont {Endr{\H{o}}czi}}, \bibinfo {author} {\bibfnamefont
  {V.}~\bibnamefont {Fafone}}, \bibinfo {author} {\bibfnamefont
  {S.}~\bibnamefont {Farinon}}, \bibinfo {author} {\bibfnamefont
  {I.}~\bibnamefont {Ferrante}}, \bibinfo {author} {\bibfnamefont
  {F.}~\bibnamefont {Ferrini}}, \bibinfo {author} {\bibfnamefont
  {F.}~\bibnamefont {Fidecaro}}, \bibinfo {author} {\bibfnamefont
  {I.}~\bibnamefont {Fiori}}, \bibinfo {author} {\bibfnamefont
  {R.}~\bibnamefont {Flaminio}}, \bibinfo {author} {\bibfnamefont {J.-D.}\
  \bibnamefont {Fournier}}, \bibinfo {author} {\bibfnamefont {S.}~\bibnamefont
  {Franco}}, \bibinfo {author} {\bibfnamefont {S.}~\bibnamefont {Frasca}},
  \bibinfo {author} {\bibfnamefont {F.}~\bibnamefont {Frasconi}}, \bibinfo
  {author} {\bibfnamefont {L.}~\bibnamefont {Gammaitoni}}, \bibinfo {author}
  {\bibfnamefont {F.}~\bibnamefont {Garufi}}, \bibinfo {author} {\bibfnamefont
  {M.}~\bibnamefont {Gaspard}}, \bibinfo {author} {\bibfnamefont
  {A.}~\bibnamefont {Gatto}}, \bibinfo {author} {\bibfnamefont
  {G.}~\bibnamefont {Gemme}}, \bibinfo {author} {\bibfnamefont
  {B.}~\bibnamefont {Gendre}}, \bibinfo {author} {\bibfnamefont
  {E.}~\bibnamefont {Genin}}, \bibinfo {author} {\bibfnamefont
  {A.}~\bibnamefont {Gennai}}, \bibinfo {author} {\bibfnamefont
  {S.}~\bibnamefont {Ghosh}}, \bibinfo {author} {\bibfnamefont
  {L.}~\bibnamefont {Giacobone}}, \bibinfo {author} {\bibfnamefont
  {A.}~\bibnamefont {Giazotto}}, \bibinfo {author} {\bibfnamefont
  {R.}~\bibnamefont {Gouaty}}, \bibinfo {author} {\bibfnamefont
  {M.}~\bibnamefont {Granata}}, \bibinfo {author} {\bibfnamefont
  {G.}~\bibnamefont {Greco}}, \bibinfo {author} {\bibfnamefont
  {P.}~\bibnamefont {Groot}}, \bibinfo {author} {\bibfnamefont {G.~M.}\
  \bibnamefont {Guidi}}, \bibinfo {author} {\bibfnamefont {J.}~\bibnamefont
  {Harms}}, \bibinfo {author} {\bibfnamefont {A.}~\bibnamefont {Heidmann}},
  \bibinfo {author} {\bibfnamefont {H.}~\bibnamefont {Heitmann}}, \bibinfo
  {author} {\bibfnamefont {P.}~\bibnamefont {Hello}}, \bibinfo {author}
  {\bibfnamefont {G.}~\bibnamefont {Hemming}}, \bibinfo {author} {\bibfnamefont
  {E.}~\bibnamefont {Hennes}}, \bibinfo {author} {\bibfnamefont
  {D.}~\bibnamefont {Hofman}}, \bibinfo {author} {\bibfnamefont
  {P.}~\bibnamefont {Jaranowski}}, \bibinfo {author} {\bibfnamefont {R.~J.~G.}\
  \bibnamefont {Jonker}}, \bibinfo {author} {\bibfnamefont {M.}~\bibnamefont
  {Kasprzack}}, \bibinfo {author} {\bibfnamefont {F.}~\bibnamefont
  {K{\'{e}}f{\'{e}}lian}}, \bibinfo {author} {\bibfnamefont {I.}~\bibnamefont
  {Kowalska}}, \bibinfo {author} {\bibfnamefont {M.}~\bibnamefont {Kraan}},
  \bibinfo {author} {\bibfnamefont {A.}~\bibnamefont {Kr{\'{o}}lak}}, \bibinfo
  {author} {\bibfnamefont {A.}~\bibnamefont {Kutynia}}, \bibinfo {author}
  {\bibfnamefont {C.}~\bibnamefont {Lazzaro}}, \bibinfo {author} {\bibfnamefont
  {M.}~\bibnamefont {Leonardi}}, \bibinfo {author} {\bibfnamefont
  {N.}~\bibnamefont {Leroy}}, \bibinfo {author} {\bibfnamefont
  {N.}~\bibnamefont {Letendre}}, \bibinfo {author} {\bibfnamefont {T.~G.~F.}\
  \bibnamefont {Li}}, \bibinfo {author} {\bibfnamefont {B.}~\bibnamefont
  {Lieunard}}, \bibinfo {author} {\bibfnamefont {M.}~\bibnamefont {Lorenzini}},
  \bibinfo {author} {\bibfnamefont {V.}~\bibnamefont {Loriette}}, \bibinfo
  {author} {\bibfnamefont {G.}~\bibnamefont {Losurdo}}, \bibinfo {author}
  {\bibfnamefont {C.}~\bibnamefont {Magazz{\`{u}}}}, \bibinfo {author}
  {\bibfnamefont {E.}~\bibnamefont {Majorana}}, \bibinfo {author}
  {\bibfnamefont {I.}~\bibnamefont {Maksimovic}}, \bibinfo {author}
  {\bibfnamefont {V.}~\bibnamefont {Malvezzi}}, \bibinfo {author}
  {\bibfnamefont {N.}~\bibnamefont {Man}}, \bibinfo {author} {\bibfnamefont
  {V.}~\bibnamefont {Mangano}}, \bibinfo {author} {\bibfnamefont
  {M.}~\bibnamefont {Mantovani}}, \bibinfo {author} {\bibfnamefont
  {F.}~\bibnamefont {Marchesoni}}, \bibinfo {author} {\bibfnamefont
  {F.}~\bibnamefont {Marion}}, \bibinfo {author} {\bibfnamefont
  {J.}~\bibnamefont {Marque}}, \bibinfo {author} {\bibfnamefont
  {F.}~\bibnamefont {Martelli}}, \bibinfo {author} {\bibfnamefont
  {L.}~\bibnamefont {Martellini}}, \bibinfo {author} {\bibfnamefont
  {A.}~\bibnamefont {Masserot}}, \bibinfo {author} {\bibfnamefont
  {D.}~\bibnamefont {Meacher}}, \bibinfo {author} {\bibfnamefont
  {J.}~\bibnamefont {Meidam}}, \bibinfo {author} {\bibfnamefont
  {F.}~\bibnamefont {Mezzani}}, \bibinfo {author} {\bibfnamefont
  {C.}~\bibnamefont {Michel}}, \bibinfo {author} {\bibfnamefont
  {L.}~\bibnamefont {Milano}}, \bibinfo {author} {\bibfnamefont
  {Y.}~\bibnamefont {Minenkov}}, \bibinfo {author} {\bibfnamefont
  {A.}~\bibnamefont {Moggi}}, \bibinfo {author} {\bibfnamefont
  {M.}~\bibnamefont {Mohan}}, \bibinfo {author} {\bibfnamefont
  {M.}~\bibnamefont {Montani}}, \bibinfo {author} {\bibfnamefont
  {N.}~\bibnamefont {Morgado}}, \bibinfo {author} {\bibfnamefont
  {B.}~\bibnamefont {Mours}}, \bibinfo {author} {\bibfnamefont
  {F.}~\bibnamefont {Mul}}, \bibinfo {author} {\bibfnamefont {M.~F.}\
  \bibnamefont {Nagy}}, \bibinfo {author} {\bibfnamefont {I.}~\bibnamefont
  {Nardecchia}}, \bibinfo {author} {\bibfnamefont {L.}~\bibnamefont
  {Naticchioni}}, \bibinfo {author} {\bibfnamefont {G.}~\bibnamefont
  {Nelemans}}, \bibinfo {author} {\bibfnamefont {I.}~\bibnamefont {Neri}},
  \bibinfo {author} {\bibfnamefont {M.}~\bibnamefont {Neri}}, \bibinfo {author}
  {\bibfnamefont {F.}~\bibnamefont {Nocera}}, \bibinfo {author} {\bibfnamefont
  {E.}~\bibnamefont {Pacaud}}, \bibinfo {author} {\bibfnamefont
  {C.}~\bibnamefont {Palomba}}, \bibinfo {author} {\bibfnamefont
  {F.}~\bibnamefont {Paoletti}}, \bibinfo {author} {\bibfnamefont
  {A.}~\bibnamefont {Paoli}}, \bibinfo {author} {\bibfnamefont
  {A.}~\bibnamefont {Pasqualetti}}, \bibinfo {author} {\bibfnamefont
  {R.}~\bibnamefont {Passaquieti}}, \bibinfo {author} {\bibfnamefont
  {D.}~\bibnamefont {Passuello}}, \bibinfo {author} {\bibfnamefont
  {M.}~\bibnamefont {Perciballi}}, \bibinfo {author} {\bibfnamefont
  {S.}~\bibnamefont {Petit}}, \bibinfo {author} {\bibfnamefont
  {M.}~\bibnamefont {Pichot}}, \bibinfo {author} {\bibfnamefont
  {F.}~\bibnamefont {Piergiovanni}}, \bibinfo {author} {\bibfnamefont
  {G.}~\bibnamefont {Pillant}}, \bibinfo {author} {\bibfnamefont
  {A.}~\bibnamefont {Piluso}}, \bibinfo {author} {\bibfnamefont
  {L.}~\bibnamefont {Pinard}}, \bibinfo {author} {\bibfnamefont
  {R.}~\bibnamefont {Poggiani}}, \bibinfo {author} {\bibfnamefont
  {M.}~\bibnamefont {Prijatelj}}, \bibinfo {author} {\bibfnamefont {G.~A.}\
  \bibnamefont {Prodi}}, \bibinfo {author} {\bibfnamefont {M.}~\bibnamefont
  {Punturo}}, \bibinfo {author} {\bibfnamefont {P.}~\bibnamefont {Puppo}},
  \bibinfo {author} {\bibfnamefont {D.~S.}\ \bibnamefont {Rabeling}}, \bibinfo
  {author} {\bibfnamefont {I.}~\bibnamefont {R{\'{a}}cz}}, \bibinfo {author}
  {\bibfnamefont {P.}~\bibnamefont {Rapagnani}}, \bibinfo {author}
  {\bibfnamefont {M.}~\bibnamefont {Razzano}}, \bibinfo {author} {\bibfnamefont
  {V.}~\bibnamefont {Re}}, \bibinfo {author} {\bibfnamefont {T.}~\bibnamefont
  {Regimbau}}, \bibinfo {author} {\bibfnamefont {F.}~\bibnamefont {Ricci}},
  \bibinfo {author} {\bibfnamefont {F.}~\bibnamefont {Robinet}}, \bibinfo
  {author} {\bibfnamefont {A.}~\bibnamefont {Rocchi}}, \bibinfo {author}
  {\bibfnamefont {L.}~\bibnamefont {Rolland}}, \bibinfo {author} {\bibfnamefont
  {R.}~\bibnamefont {Romano}}, \bibinfo {author} {\bibfnamefont
  {D.}~\bibnamefont {Rosi{\'{n}}ska}}, \bibinfo {author} {\bibfnamefont
  {P.}~\bibnamefont {Ruggi}}, \bibinfo {author} {\bibfnamefont
  {E.}~\bibnamefont {Saracco}}, \bibinfo {author} {\bibfnamefont
  {B.}~\bibnamefont {Sassolas}}, \bibinfo {author} {\bibfnamefont
  {F.}~\bibnamefont {Schimmel}}, \bibinfo {author} {\bibfnamefont
  {D.}~\bibnamefont {Sentenac}}, \bibinfo {author} {\bibfnamefont
  {V.}~\bibnamefont {Sequino}}, \bibinfo {author} {\bibfnamefont
  {S.}~\bibnamefont {Shah}}, \bibinfo {author} {\bibfnamefont {K.}~\bibnamefont
  {Siellez}}, \bibinfo {author} {\bibfnamefont {N.}~\bibnamefont {Straniero}},
  \bibinfo {author} {\bibfnamefont {B.}~\bibnamefont {Swinkels}}, \bibinfo
  {author} {\bibfnamefont {M.}~\bibnamefont {Tacca}}, \bibinfo {author}
  {\bibfnamefont {M.}~\bibnamefont {Tonelli}}, \bibinfo {author} {\bibfnamefont
  {F.}~\bibnamefont {Travasso}}, \bibinfo {author} {\bibfnamefont
  {M.}~\bibnamefont {Turconi}}, \bibinfo {author} {\bibfnamefont
  {G.}~\bibnamefont {Vajente}}, \bibinfo {author} {\bibfnamefont
  {N.}~\bibnamefont {van Bakel}}, \bibinfo {author} {\bibfnamefont
  {M.}~\bibnamefont {van Beuzekom}}, \bibinfo {author} {\bibfnamefont
  {J.~F.~J.}\ \bibnamefont {van~den Brand}}, \bibinfo {author} {\bibfnamefont
  {C.~V.~D.}\ \bibnamefont {Broeck}}, \bibinfo {author} {\bibfnamefont {M.~V.}\
  \bibnamefont {van~der Sluys}}, \bibinfo {author} {\bibfnamefont
  {J.}~\bibnamefont {van Heijningen}}, \bibinfo {author} {\bibfnamefont
  {M.}~\bibnamefont {Vas{\'{u}}th}}, \bibinfo {author} {\bibfnamefont
  {G.}~\bibnamefont {Vedovato}}, \bibinfo {author} {\bibfnamefont
  {J.}~\bibnamefont {Veitch}}, \bibinfo {author} {\bibfnamefont
  {D.}~\bibnamefont {Verkindt}}, \bibinfo {author} {\bibfnamefont
  {F.}~\bibnamefont {Vetrano}}, \bibinfo {author} {\bibfnamefont
  {A.}~\bibnamefont {Vicer{\'{e}}}}, \bibinfo {author} {\bibfnamefont {J.-Y.}\
  \bibnamefont {Vinet}}, \bibinfo {author} {\bibfnamefont {G.}~\bibnamefont
  {Visser}}, \bibinfo {author} {\bibfnamefont {H.}~\bibnamefont {Vocca}},
  \bibinfo {author} {\bibfnamefont {R.}~\bibnamefont {Ward}}, \bibinfo {author}
  {\bibfnamefont {M.}~\bibnamefont {Was}}, \bibinfo {author} {\bibfnamefont
  {L.-W.}\ \bibnamefont {Wei}}, \bibinfo {author} {\bibfnamefont
  {M.}~\bibnamefont {Yvert}}, \bibinfo {author} {\bibfnamefont {A.~Z.}\
  \bibnamefont {{\.{z}}ny}}, \ and\ \bibinfo {author} {\bibfnamefont {J.-P.}\
  \bibnamefont {Zendri}},\ }\href {\doibase 10.1088/0264-9381/32/2/024001}
  {\bibfield  {journal} {\bibinfo  {journal} {Classical and Quantum Gravity}\
  }\textbf {\bibinfo {volume} {32}},\ \bibinfo {pages} {024001} (\bibinfo
  {year} {2014})}\BibitemShut {NoStop}%
\bibitem [{\citenamefont {Dooley}\ and\ \citenamefont {(for~the LIGO
  Scientific~Collaboration)}(2015)}]{Dooley_2015}%
  \BibitemOpen
  \bibfield  {author} {\bibinfo {author} {\bibfnamefont {K.~L.}\ \bibnamefont
  {Dooley}}\ and\ \bibinfo {author} {\bibnamefont {(for~the LIGO
  Scientific~Collaboration)}},\ }\href {\doibase
  10.1088/1742-6596/610/1/012015} {\bibfield  {journal} {\bibinfo  {journal}
  {Journal of Physics: Conference Series}\ }\textbf {\bibinfo {volume} {610}},\
  \bibinfo {pages} {012015} (\bibinfo {year} {2015})}\BibitemShut {NoStop}%
\bibitem [{\citenamefont {Chickarmane}\ \emph {et~al.}(1998)\citenamefont
  {Chickarmane}, \citenamefont {Dhurandhar}, \citenamefont {Ralph},
  \citenamefont {Gray}, \citenamefont {Bachor},\ and\ \citenamefont
  {McClelland}}]{PhysRevA.57.3898}%
  \BibitemOpen
  \bibfield  {author} {\bibinfo {author} {\bibfnamefont {V.}~\bibnamefont
  {Chickarmane}}, \bibinfo {author} {\bibfnamefont {S.~V.}\ \bibnamefont
  {Dhurandhar}}, \bibinfo {author} {\bibfnamefont {T.~C.}\ \bibnamefont
  {Ralph}}, \bibinfo {author} {\bibfnamefont {M.}~\bibnamefont {Gray}},
  \bibinfo {author} {\bibfnamefont {H.-A.}\ \bibnamefont {Bachor}}, \ and\
  \bibinfo {author} {\bibfnamefont {D.~E.}\ \bibnamefont {McClelland}},\ }\href
  {\doibase 10.1103/PhysRevA.57.3898} {\bibfield  {journal} {\bibinfo
  {journal} {Phys. Rev. A}\ }\textbf {\bibinfo {volume} {57}},\ \bibinfo
  {pages} {3898} (\bibinfo {year} {1998})}\BibitemShut {NoStop}%
\bibitem [{\citenamefont {Rakhmanov}(2001)}]{Rakhmanov:01}%
  \BibitemOpen
  \bibfield  {author} {\bibinfo {author} {\bibfnamefont {M.}~\bibnamefont
  {Rakhmanov}},\ }\href {\doibase 10.1364/AO.40.006596} {\bibfield  {journal}
  {\bibinfo  {journal} {Appl. Opt.}\ }\textbf {\bibinfo {volume} {40}},\
  \bibinfo {pages} {6596} (\bibinfo {year} {2001})}\BibitemShut {NoStop}%
\bibitem [{\citenamefont {Ward}(2010)}]{Ward_THESIS_2010}%
  \BibitemOpen
  \bibfield  {author} {\bibinfo {author} {\bibfnamefont {R.}~\bibnamefont
  {Ward}},\ }\emph {\bibinfo {title} {Length Sensing and Control of a Prototype
  Advanced Interferometric Gravitational Wave Detector}},\ \href
  {https://thesis.library.caltech.edu/5836/} {Ph.D. thesis},\ \bibinfo
  {school} {Caltech} (\bibinfo {year} {2010})\BibitemShut {NoStop}%
\bibitem [{\citenamefont {Niebauer}\ \emph {et~al.}(1991)\citenamefont
  {Niebauer}, \citenamefont {Schilling}, \citenamefont {Danzmann},
  \citenamefont {R\"udiger},\ and\ \citenamefont {Winkler}}]{PhysRevA.43.5022}%
  \BibitemOpen
  \bibfield  {author} {\bibinfo {author} {\bibfnamefont {T.~M.}\ \bibnamefont
  {Niebauer}}, \bibinfo {author} {\bibfnamefont {R.}~\bibnamefont {Schilling}},
  \bibinfo {author} {\bibfnamefont {K.}~\bibnamefont {Danzmann}}, \bibinfo
  {author} {\bibfnamefont {A.}~\bibnamefont {R\"udiger}}, \ and\ \bibinfo
  {author} {\bibfnamefont {W.}~\bibnamefont {Winkler}},\ }\href {\doibase
  10.1103/PhysRevA.43.5022} {\bibfield  {journal} {\bibinfo  {journal} {Phys.
  Rev. A}\ }\textbf {\bibinfo {volume} {43}},\ \bibinfo {pages} {5022}
  (\bibinfo {year} {1991})}\BibitemShut {NoStop}%
\bibitem [{\citenamefont {Buonanno}\ \emph {et~al.}(2003)\citenamefont
  {Buonanno}, \citenamefont {Chen},\ and\ \citenamefont
  {Mavalvala}}]{PhysRevD.67.122005}%
  \BibitemOpen
  \bibfield  {author} {\bibinfo {author} {\bibfnamefont {A.}~\bibnamefont
  {Buonanno}}, \bibinfo {author} {\bibfnamefont {Y.}~\bibnamefont {Chen}}, \
  and\ \bibinfo {author} {\bibfnamefont {N.}~\bibnamefont {Mavalvala}},\ }\href
  {\doibase 10.1103/PhysRevD.67.122005} {\bibfield  {journal} {\bibinfo
  {journal} {Phys. Rev. D}\ }\textbf {\bibinfo {volume} {67}},\ \bibinfo
  {pages} {122005} (\bibinfo {year} {2003})}\BibitemShut {NoStop}%
\bibitem [{\citenamefont {Meers}\ and\ \citenamefont
  {Strain}(1991)}]{PhysRevA.44.4693}%
  \BibitemOpen
  \bibfield  {author} {\bibinfo {author} {\bibfnamefont {B.~J.}\ \bibnamefont
  {Meers}}\ and\ \bibinfo {author} {\bibfnamefont {K.~A.}\ \bibnamefont
  {Strain}},\ }\href {\doibase 10.1103/PhysRevA.44.4693} {\bibfield  {journal}
  {\bibinfo  {journal} {Phys. Rev. A}\ }\textbf {\bibinfo {volume} {44}},\
  \bibinfo {pages} {4693} (\bibinfo {year} {1991})}\BibitemShut {NoStop}%
\bibitem [{\citenamefont {Caves}(1981)}]{PhysRevD.23.1693}%
  \BibitemOpen
  \bibfield  {author} {\bibinfo {author} {\bibfnamefont {C.~M.}\ \bibnamefont
  {Caves}},\ }\href {\doibase 10.1103/PhysRevD.23.1693} {\bibfield  {journal}
  {\bibinfo  {journal} {Phys. Rev. D}\ }\textbf {\bibinfo {volume} {23}},\
  \bibinfo {pages} {1693} (\bibinfo {year} {1981})}\BibitemShut {NoStop}%
\bibitem [{\citenamefont {Gea-Banacloche}\ and\ \citenamefont
  {Leuchs}(1987)}]{Gea-Banacloche1987}%
  \BibitemOpen
  \bibfield  {author} {\bibinfo {author} {\bibfnamefont {J.}~\bibnamefont
  {Gea-Banacloche}}\ and\ \bibinfo {author} {\bibfnamefont {G.}~\bibnamefont
  {Leuchs}},\ }\href {\doibase 10.1080/09500348714550751} {\bibfield  {journal}
  {\bibinfo  {journal} {Journal of Modern Optics}\ }\textbf {\bibinfo {volume}
  {34}},\ \bibinfo {pages} {793} (\bibinfo {year} {1987})}\BibitemShut
  {NoStop}%
\bibitem [{\citenamefont {Fritschel}\ \emph {et~al.}(2014)\citenamefont
  {Fritschel}, \citenamefont {Evans},\ and\ \citenamefont
  {Frolov}}]{Fritschel2014}%
  \BibitemOpen
  \bibfield  {author} {\bibinfo {author} {\bibfnamefont {P.}~\bibnamefont
  {Fritschel}}, \bibinfo {author} {\bibfnamefont {M.}~\bibnamefont {Evans}}, \
  and\ \bibinfo {author} {\bibfnamefont {V.}~\bibnamefont {Frolov}},\ }\href
  {\doibase 10.1364/OE.22.004224} {\bibfield  {journal} {\bibinfo  {journal}
  {Opt. Express}\ }\textbf {\bibinfo {volume} {22}},\ \bibinfo {pages} {4224}
  (\bibinfo {year} {2014})}\BibitemShut {NoStop}%
\bibitem [{\citenamefont {Zhang}\ \emph {et~al.}(2017)\citenamefont {Zhang},
  \citenamefont {Danilishin}, \citenamefont {Steinlechner}, \citenamefont
  {Barr}, \citenamefont {Bell}, \citenamefont {Dupej}, \citenamefont {Gr\"af},
  \citenamefont {Hennig}, \citenamefont {Houston}, \citenamefont {Huttner},
  \citenamefont {Leavey}, \citenamefont {Pascucci}, \citenamefont {Sorazu},
  \citenamefont {Spencer}, \citenamefont {Wright}, \citenamefont {Strain},\
  and\ \citenamefont {Hild}}]{PhysRevD.95.062001}%
  \BibitemOpen
  \bibfield  {author} {\bibinfo {author} {\bibfnamefont {T.}~\bibnamefont
  {Zhang}}, \bibinfo {author} {\bibfnamefont {S.~L.}\ \bibnamefont
  {Danilishin}}, \bibinfo {author} {\bibfnamefont {S.}~\bibnamefont
  {Steinlechner}}, \bibinfo {author} {\bibfnamefont {B.~W.}\ \bibnamefont
  {Barr}}, \bibinfo {author} {\bibfnamefont {A.~S.}\ \bibnamefont {Bell}},
  \bibinfo {author} {\bibfnamefont {P.}~\bibnamefont {Dupej}}, \bibinfo
  {author} {\bibfnamefont {C.}~\bibnamefont {Gr\"af}}, \bibinfo {author}
  {\bibfnamefont {J.-S.}\ \bibnamefont {Hennig}}, \bibinfo {author}
  {\bibfnamefont {E.~A.}\ \bibnamefont {Houston}}, \bibinfo {author}
  {\bibfnamefont {S.~H.}\ \bibnamefont {Huttner}}, \bibinfo {author}
  {\bibfnamefont {S.~S.}\ \bibnamefont {Leavey}}, \bibinfo {author}
  {\bibfnamefont {D.}~\bibnamefont {Pascucci}}, \bibinfo {author}
  {\bibfnamefont {B.}~\bibnamefont {Sorazu}}, \bibinfo {author} {\bibfnamefont
  {A.}~\bibnamefont {Spencer}}, \bibinfo {author} {\bibfnamefont
  {J.}~\bibnamefont {Wright}}, \bibinfo {author} {\bibfnamefont {K.~A.}\
  \bibnamefont {Strain}}, \ and\ \bibinfo {author} {\bibfnamefont
  {S.}~\bibnamefont {Hild}},\ }\href {\doibase 10.1103/PhysRevD.95.062001}
  {\bibfield  {journal} {\bibinfo  {journal} {Phys. Rev. D}\ }\textbf {\bibinfo
  {volume} {95}},\ \bibinfo {pages} {062001} (\bibinfo {year}
  {2017})}\BibitemShut {NoStop}%
\bibitem [{\citenamefont {Steinlechner}\ \emph {et~al.}(2015)\citenamefont
  {Steinlechner}, \citenamefont {Barr}, \citenamefont {Bell}, \citenamefont
  {Danilishin}, \citenamefont {Gl\"afke}, \citenamefont {Gr\"af}, \citenamefont
  {Hennig}, \citenamefont {Houston}, \citenamefont {Huttner}, \citenamefont
  {Leavey}, \citenamefont {Pascucci}, \citenamefont {Sorazu}, \citenamefont
  {Spencer}, \citenamefont {Strain}, \citenamefont {Wright},\ and\
  \citenamefont {Hild}}]{Sebastian2015}%
  \BibitemOpen
  \bibfield  {author} {\bibinfo {author} {\bibfnamefont {S.}~\bibnamefont
  {Steinlechner}}, \bibinfo {author} {\bibfnamefont {B.~W.}\ \bibnamefont
  {Barr}}, \bibinfo {author} {\bibfnamefont {A.~S.}\ \bibnamefont {Bell}},
  \bibinfo {author} {\bibfnamefont {S.~L.}\ \bibnamefont {Danilishin}},
  \bibinfo {author} {\bibfnamefont {A.}~\bibnamefont {Gl\"afke}}, \bibinfo
  {author} {\bibfnamefont {C.}~\bibnamefont {Gr\"af}}, \bibinfo {author}
  {\bibfnamefont {J.-S.}\ \bibnamefont {Hennig}}, \bibinfo {author}
  {\bibfnamefont {E.~A.}\ \bibnamefont {Houston}}, \bibinfo {author}
  {\bibfnamefont {S.~H.}\ \bibnamefont {Huttner}}, \bibinfo {author}
  {\bibfnamefont {S.~S.}\ \bibnamefont {Leavey}}, \bibinfo {author}
  {\bibfnamefont {D.}~\bibnamefont {Pascucci}}, \bibinfo {author}
  {\bibfnamefont {B.}~\bibnamefont {Sorazu}}, \bibinfo {author} {\bibfnamefont
  {A.}~\bibnamefont {Spencer}}, \bibinfo {author} {\bibfnamefont {K.~A.}\
  \bibnamefont {Strain}}, \bibinfo {author} {\bibfnamefont {J.}~\bibnamefont
  {Wright}}, \ and\ \bibinfo {author} {\bibfnamefont {S.}~\bibnamefont
  {Hild}},\ }\href {\doibase 10.1103/PhysRevD.92.072009} {\bibfield  {journal}
  {\bibinfo  {journal} {Phys. Rev. D}\ }\textbf {\bibinfo {volume} {92}},\
  \bibinfo {pages} {072009} (\bibinfo {year} {2015})}\BibitemShut {NoStop}%
\bibitem [{\citenamefont {Zhang}\ \emph {et~al.}(2018)\citenamefont {Zhang},
  \citenamefont {Knyazev}, \citenamefont {Steinlechner}, \citenamefont
  {Khalili}, \citenamefont {Barr}, \citenamefont {Bell}, \citenamefont {Dupej},
  \citenamefont {Briggs}, \citenamefont {Gr{\~A}€f}, \citenamefont
  {Callaghan}, \citenamefont {Hennig}, \citenamefont {Houston}, \citenamefont
  {Huttner}, \citenamefont {Leavey}, \citenamefont {Pascucci}, \citenamefont
  {Sorazu}, \citenamefont {Spencer}, \citenamefont {Wright}, \citenamefont
  {Strain}, \citenamefont {Hild},\ and\ \citenamefont
  {Danilishin}}]{Zhang_2018}%
  \BibitemOpen
  \bibfield  {author} {\bibinfo {author} {\bibfnamefont {T.}~\bibnamefont
  {Zhang}}, \bibinfo {author} {\bibfnamefont {E.}~\bibnamefont {Knyazev}},
  \bibinfo {author} {\bibfnamefont {S.}~\bibnamefont {Steinlechner}}, \bibinfo
  {author} {\bibfnamefont {F.~Y.}\ \bibnamefont {Khalili}}, \bibinfo {author}
  {\bibfnamefont {B.~W.}\ \bibnamefont {Barr}}, \bibinfo {author}
  {\bibfnamefont {A.~S.}\ \bibnamefont {Bell}}, \bibinfo {author}
  {\bibfnamefont {P.}~\bibnamefont {Dupej}}, \bibinfo {author} {\bibfnamefont
  {J.}~\bibnamefont {Briggs}}, \bibinfo {author} {\bibfnamefont
  {C.}~\bibnamefont {Gr{\~A}€f}}, \bibinfo {author} {\bibfnamefont
  {J.}~\bibnamefont {Callaghan}}, \bibinfo {author} {\bibfnamefont {J.~S.}\
  \bibnamefont {Hennig}}, \bibinfo {author} {\bibfnamefont {E.~A.}\
  \bibnamefont {Houston}}, \bibinfo {author} {\bibfnamefont {S.~H.}\
  \bibnamefont {Huttner}}, \bibinfo {author} {\bibfnamefont {S.~S.}\
  \bibnamefont {Leavey}}, \bibinfo {author} {\bibfnamefont {D.}~\bibnamefont
  {Pascucci}}, \bibinfo {author} {\bibfnamefont {B.}~\bibnamefont {Sorazu}},
  \bibinfo {author} {\bibfnamefont {A.}~\bibnamefont {Spencer}}, \bibinfo
  {author} {\bibfnamefont {J.}~\bibnamefont {Wright}}, \bibinfo {author}
  {\bibfnamefont {K.~A.}\ \bibnamefont {Strain}}, \bibinfo {author}
  {\bibfnamefont {S.}~\bibnamefont {Hild}}, \ and\ \bibinfo {author}
  {\bibfnamefont {S.~L.}\ \bibnamefont {Danilishin}},\ }\href {\doibase
  10.1088/1367-2630/aae86e} {\bibfield  {journal} {\bibinfo  {journal} {New
  Journal of Physics}\ }\textbf {\bibinfo {volume} {20}},\ \bibinfo {pages}
  {103040} (\bibinfo {year} {2018})}\BibitemShut {NoStop}%
\bibitem [{\citenamefont {Aasi}\ \emph
  {et~al.}(2015{\natexlab{b}})\citenamefont {Aasi}, \citenamefont {Abbott},
  \citenamefont {Abbott}, \citenamefont {Abbott} \emph {et~al.}}]{2015}%
  \BibitemOpen
  \bibfield  {author} {\bibinfo {author} {\bibfnamefont {J.}~\bibnamefont
  {Aasi}}, \bibinfo {author} {\bibfnamefont {B.~P.}\ \bibnamefont {Abbott}},
  \bibinfo {author} {\bibfnamefont {R.}~\bibnamefont {Abbott}}, \bibinfo
  {author} {\bibfnamefont {T.}~\bibnamefont {Abbott}},  \emph {et~al.},\ }\href
  {\doibase 10.1088/0264-9381/32/7/074001} {\bibfield  {journal} {\bibinfo
  {journal} {Classical and Quantum Gravity}\ }\textbf {\bibinfo {volume}
  {32}},\ \bibinfo {pages} {074001} (\bibinfo {year}
  {2015}{\natexlab{b}})}\BibitemShut {NoStop}%
\bibitem [{\citenamefont {Hild}\ \emph {et~al.}(2009)\citenamefont {Hild},
  \citenamefont {Grote}, \citenamefont {Degallaix}, \citenamefont {Chelkowski},
  \citenamefont {Danzmann}, \citenamefont {Freise}, \citenamefont {Hewitson},
  \citenamefont {Hough}, \citenamefont {Lück}, \citenamefont {Prijatelj},
  \citenamefont {Strain}, \citenamefont {Smith},\ and\ \citenamefont
  {Willke}}]{Hild_2009}%
  \BibitemOpen
  \bibfield  {author} {\bibinfo {author} {\bibfnamefont {S.}~\bibnamefont
  {Hild}}, \bibinfo {author} {\bibfnamefont {H.}~\bibnamefont {Grote}},
  \bibinfo {author} {\bibfnamefont {J.}~\bibnamefont {Degallaix}}, \bibinfo
  {author} {\bibfnamefont {S.}~\bibnamefont {Chelkowski}}, \bibinfo {author}
  {\bibfnamefont {K.}~\bibnamefont {Danzmann}}, \bibinfo {author}
  {\bibfnamefont {A.}~\bibnamefont {Freise}}, \bibinfo {author} {\bibfnamefont
  {M.}~\bibnamefont {Hewitson}}, \bibinfo {author} {\bibfnamefont
  {J.}~\bibnamefont {Hough}}, \bibinfo {author} {\bibfnamefont
  {H.}~\bibnamefont {Lück}}, \bibinfo {author} {\bibfnamefont
  {M.}~\bibnamefont {Prijatelj}}, \bibinfo {author} {\bibfnamefont {K.~A.}\
  \bibnamefont {Strain}}, \bibinfo {author} {\bibfnamefont {J.~R.}\
  \bibnamefont {Smith}}, \ and\ \bibinfo {author} {\bibfnamefont
  {B.}~\bibnamefont {Willke}},\ }\href {\doibase 10.1088/0264-9381/26/5/055012}
  {\bibfield  {journal} {\bibinfo  {journal} {Classical and Quantum Gravity}\
  }\textbf {\bibinfo {volume} {26}},\ \bibinfo {pages} {055012} (\bibinfo
  {year} {2009})}\BibitemShut {NoStop}%
\bibitem [{\citenamefont {Fricke}\ \emph {et~al.}(2012)\citenamefont {Fricke},
  \citenamefont {Smith-Lefebvre}, \citenamefont {Abbott}, \citenamefont
  {Adhikari}, \citenamefont {Dooley}, \citenamefont {Evans}, \citenamefont
  {Fritschel}, \citenamefont {Frolov}, \citenamefont {Kawabe}, \citenamefont
  {Kissel}, \citenamefont {Slagmolen},\ and\ \citenamefont
  {Waldman}}]{Fricke_2012}%
  \BibitemOpen
  \bibfield  {author} {\bibinfo {author} {\bibfnamefont {T.~T.}\ \bibnamefont
  {Fricke}}, \bibinfo {author} {\bibfnamefont {N.~D.}\ \bibnamefont
  {Smith-Lefebvre}}, \bibinfo {author} {\bibfnamefont {R.}~\bibnamefont
  {Abbott}}, \bibinfo {author} {\bibfnamefont {R.}~\bibnamefont {Adhikari}},
  \bibinfo {author} {\bibfnamefont {K.~L.}\ \bibnamefont {Dooley}}, \bibinfo
  {author} {\bibfnamefont {M.}~\bibnamefont {Evans}}, \bibinfo {author}
  {\bibfnamefont {P.}~\bibnamefont {Fritschel}}, \bibinfo {author}
  {\bibfnamefont {V.~V.}\ \bibnamefont {Frolov}}, \bibinfo {author}
  {\bibfnamefont {K.}~\bibnamefont {Kawabe}}, \bibinfo {author} {\bibfnamefont
  {J.~S.}\ \bibnamefont {Kissel}}, \bibinfo {author} {\bibfnamefont {B.~J.~J.}\
  \bibnamefont {Slagmolen}}, \ and\ \bibinfo {author} {\bibfnamefont {S.~J.}\
  \bibnamefont {Waldman}},\ }\href {\doibase 10.1088/0264-9381/29/6/065005}
  {\bibfield  {journal} {\bibinfo  {journal} {Classical and Quantum Gravity}\
  }\textbf {\bibinfo {volume} {29}},\ \bibinfo {pages} {065005} (\bibinfo
  {year} {2012})}\BibitemShut {NoStop}%
\bibitem [{\citenamefont {and}(2010)}]{Grote_2010}%
  \BibitemOpen
  \bibfield  {author} {\bibinfo {author} {\bibfnamefont {H.~G.}\ \bibnamefont
  {and}},\ }\href {\doibase 10.1088/0264-9381/27/8/084003} {\bibfield
  {journal} {\bibinfo  {journal} {Classical and Quantum Gravity}\ }\textbf
  {\bibinfo {volume} {27}},\ \bibinfo {pages} {084003} (\bibinfo {year}
  {2010})}\BibitemShut {NoStop}%
\bibitem [{\citenamefont {Akutsu}\ \emph {et~al.}(2019)\citenamefont {Akutsu},
  \citenamefont {Ando}, \citenamefont {Arai}, \citenamefont {Arai},
  \citenamefont {Araki}, \citenamefont {Araya},\ and\ \citenamefont
  {collaboration}}]{Akutsu:2019aa}%
  \BibitemOpen
  \bibfield  {author} {\bibinfo {author} {\bibfnamefont {T.}~\bibnamefont
  {Akutsu}}, \bibinfo {author} {\bibfnamefont {M.}~\bibnamefont {Ando}},
  \bibinfo {author} {\bibfnamefont {K.}~\bibnamefont {Arai}}, \bibinfo {author}
  {\bibfnamefont {Y.}~\bibnamefont {Arai}}, \bibinfo {author} {\bibfnamefont
  {S.}~\bibnamefont {Araki}}, \bibinfo {author} {\bibfnamefont
  {A.}~\bibnamefont {Araya}}, \ and\ \bibinfo {author} {\bibfnamefont
  {K.}~\bibnamefont {collaboration}},\ }\href {\doibase
  10.1038/s41550-018-0658-y} {\bibfield  {journal} {\bibinfo  {journal} {Nature
  Astronomy}\ }\textbf {\bibinfo {volume} {3}},\ \bibinfo {pages} {35}
  (\bibinfo {year} {2019})}\BibitemShut {NoStop}%
\bibitem [{\citenamefont {Abbott}\ \emph {et~al.}(2016)\citenamefont {Abbott},
  \citenamefont {Abbott}, \citenamefont {Abbott}, \citenamefont {Abernathy},
  \citenamefont {Acernese}, \citenamefont {Ackley}, \citenamefont {Adams},
  \citenamefont {Adams}, \citenamefont {Addesso}, \citenamefont {Adhikari}
  \emph {et~al.}}]{GW150914}%
  \BibitemOpen
  \bibfield  {author} {\bibinfo {author} {\bibfnamefont {B.}~\bibnamefont
  {Abbott}}, \bibinfo {author} {\bibfnamefont {R.}~\bibnamefont {Abbott}},
  \bibinfo {author} {\bibfnamefont {T.}~\bibnamefont {Abbott}}, \bibinfo
  {author} {\bibfnamefont {M.}~\bibnamefont {Abernathy}}, \bibinfo {author}
  {\bibfnamefont {F.}~\bibnamefont {Acernese}}, \bibinfo {author}
  {\bibfnamefont {K.}~\bibnamefont {Ackley}}, \bibinfo {author} {\bibfnamefont
  {C.}~\bibnamefont {Adams}}, \bibinfo {author} {\bibfnamefont
  {T.}~\bibnamefont {Adams}}, \bibinfo {author} {\bibfnamefont
  {P.}~\bibnamefont {Addesso}}, \bibinfo {author} {\bibfnamefont
  {R.}~\bibnamefont {Adhikari}},  \emph {et~al.} (\bibinfo {collaboration}
  {LIGO Scientific Collaboration and Virgo Collaboration}),\ }\href {\doibase
  10.1103/PhysRevLett.116.061102} {\bibfield  {journal} {\bibinfo  {journal}
  {Phys. Rev. Lett.}\ }\textbf {\bibinfo {volume} {116}},\ \bibinfo {pages}
  {061102} (\bibinfo {year} {2016})}\BibitemShut {NoStop}%
\bibitem [{\citenamefont {Abbott}\ \emph {et~al.}(2017)\citenamefont {Abbott},
  \citenamefont {Abbott}, \citenamefont {Abbott}, \citenamefont {Abernathy},
  \citenamefont {Acernese}, \citenamefont {Ackley}, \citenamefont {Adams},
  \citenamefont {Adams}, \citenamefont {Addesso}, \citenamefont {Adhikari}
  \emph {et~al.}}]{GW170817}%
  \BibitemOpen
  \bibfield  {author} {\bibinfo {author} {\bibfnamefont {B.}~\bibnamefont
  {Abbott}}, \bibinfo {author} {\bibfnamefont {R.}~\bibnamefont {Abbott}},
  \bibinfo {author} {\bibfnamefont {T.}~\bibnamefont {Abbott}}, \bibinfo
  {author} {\bibfnamefont {M.}~\bibnamefont {Abernathy}}, \bibinfo {author}
  {\bibfnamefont {F.}~\bibnamefont {Acernese}}, \bibinfo {author}
  {\bibfnamefont {K.}~\bibnamefont {Ackley}}, \bibinfo {author} {\bibfnamefont
  {C.}~\bibnamefont {Adams}}, \bibinfo {author} {\bibfnamefont
  {T.}~\bibnamefont {Adams}}, \bibinfo {author} {\bibfnamefont
  {P.}~\bibnamefont {Addesso}}, \bibinfo {author} {\bibfnamefont
  {R.}~\bibnamefont {Adhikari}},  \emph {et~al.} (\bibinfo {collaboration}
  {LIGO Scientific Collaboration and Virgo Collaboration}),\ }\href {\doibase
  10.1103/PhysRevLett.119.161101} {\bibfield  {journal} {\bibinfo  {journal}
  {Phys. Rev. Lett.}\ }\textbf {\bibinfo {volume} {119}},\ \bibinfo {pages}
  {161101} (\bibinfo {year} {2017})}\BibitemShut {NoStop}%
\bibitem [{\citenamefont {Abbott}\ \emph {et~al.}(2019)\citenamefont {Abbott},
  \citenamefont {Abbott}, \citenamefont {Abbott}, \citenamefont {Abernathy},
  \citenamefont {Acernese}, \citenamefont {Ackley}, \citenamefont {Adams},
  \citenamefont {Adams}, \citenamefont {Addesso}, \citenamefont {Adhikari}
  \emph {et~al.}}]{PhysRevX.9.031040}%
  \BibitemOpen
  \bibfield  {author} {\bibinfo {author} {\bibfnamefont {B.}~\bibnamefont
  {Abbott}}, \bibinfo {author} {\bibfnamefont {R.}~\bibnamefont {Abbott}},
  \bibinfo {author} {\bibfnamefont {T.}~\bibnamefont {Abbott}}, \bibinfo
  {author} {\bibfnamefont {M.}~\bibnamefont {Abernathy}}, \bibinfo {author}
  {\bibfnamefont {F.}~\bibnamefont {Acernese}}, \bibinfo {author}
  {\bibfnamefont {K.}~\bibnamefont {Ackley}}, \bibinfo {author} {\bibfnamefont
  {C.}~\bibnamefont {Adams}}, \bibinfo {author} {\bibfnamefont
  {T.}~\bibnamefont {Adams}}, \bibinfo {author} {\bibfnamefont
  {P.}~\bibnamefont {Addesso}}, \bibinfo {author} {\bibfnamefont
  {R.}~\bibnamefont {Adhikari}},  \emph {et~al.} (\bibinfo {collaboration}
  {LIGO Scientific Collaboration and Virgo Collaboration}),\ }\href {\doibase
  10.1103/PhysRevX.9.031040} {\bibfield  {journal} {\bibinfo  {journal} {Phys.
  Rev. X}\ }\textbf {\bibinfo {volume} {9}},\ \bibinfo {pages} {031040}
  (\bibinfo {year} {2019})}\BibitemShut {NoStop}%
\bibitem [{\citenamefont {Yu}\ \emph {et~al.}(2018)\citenamefont {Yu},
  \citenamefont {Martynov}, \citenamefont {Vitale}, \citenamefont {Evans},
  \citenamefont {Shoemaker}, \citenamefont {Barr}, \citenamefont {Hammond},
  \citenamefont {Hild}, \citenamefont {Hough}, \citenamefont {Huttner},
  \citenamefont {Rowan}, \citenamefont {Sorazu}, \citenamefont {Carbone},
  \citenamefont {Freise}, \citenamefont {Mow-Lowry}, \citenamefont {Dooley},
  \citenamefont {Fulda}, \citenamefont {Grote},\ and\ \citenamefont
  {Sigg}}]{PhysRevLett.120.141102}%
  \BibitemOpen
  \bibfield  {author} {\bibinfo {author} {\bibfnamefont {H.}~\bibnamefont
  {Yu}}, \bibinfo {author} {\bibfnamefont {D.}~\bibnamefont {Martynov}},
  \bibinfo {author} {\bibfnamefont {S.}~\bibnamefont {Vitale}}, \bibinfo
  {author} {\bibfnamefont {M.}~\bibnamefont {Evans}}, \bibinfo {author}
  {\bibfnamefont {D.}~\bibnamefont {Shoemaker}}, \bibinfo {author}
  {\bibfnamefont {B.}~\bibnamefont {Barr}}, \bibinfo {author} {\bibfnamefont
  {G.}~\bibnamefont {Hammond}}, \bibinfo {author} {\bibfnamefont
  {S.}~\bibnamefont {Hild}}, \bibinfo {author} {\bibfnamefont {J.}~\bibnamefont
  {Hough}}, \bibinfo {author} {\bibfnamefont {S.}~\bibnamefont {Huttner}},
  \bibinfo {author} {\bibfnamefont {S.}~\bibnamefont {Rowan}}, \bibinfo
  {author} {\bibfnamefont {B.}~\bibnamefont {Sorazu}}, \bibinfo {author}
  {\bibfnamefont {L.}~\bibnamefont {Carbone}}, \bibinfo {author} {\bibfnamefont
  {A.}~\bibnamefont {Freise}}, \bibinfo {author} {\bibfnamefont
  {C.}~\bibnamefont {Mow-Lowry}}, \bibinfo {author} {\bibfnamefont {K.~L.}\
  \bibnamefont {Dooley}}, \bibinfo {author} {\bibfnamefont {P.}~\bibnamefont
  {Fulda}}, \bibinfo {author} {\bibfnamefont {H.}~\bibnamefont {Grote}}, \ and\
  \bibinfo {author} {\bibfnamefont {D.}~\bibnamefont {Sigg}},\ }\href {\doibase
  10.1103/PhysRevLett.120.141102} {\bibfield  {journal} {\bibinfo  {journal}
  {Phys. Rev. Lett.}\ }\textbf {\bibinfo {volume} {120}},\ \bibinfo {pages}
  {141102} (\bibinfo {year} {2018})}\BibitemShut {NoStop}%
\bibitem [{\citenamefont {Arain}\ and\ \citenamefont
  {Mueller}(2008)}]{Arain_RECYCLING_2008}%
  \BibitemOpen
  \bibfield  {author} {\bibinfo {author} {\bibfnamefont {M.~A.}\ \bibnamefont
  {Arain}}\ and\ \bibinfo {author} {\bibfnamefont {G.}~\bibnamefont
  {Mueller}},\ }\href {\doibase 10.1364/OE.16.010018} {\bibfield  {journal}
  {\bibinfo  {journal} {Optics Express}\ }\textbf {\bibinfo {volume} {16}},\
  \bibinfo {pages} {10018} (\bibinfo {year} {2008})}\BibitemShut {NoStop}%
\bibitem [{\citenamefont {Gretarsson}\ \emph {et~al.}(2007)\citenamefont
  {Gretarsson}, \citenamefont {D'Ambrosio}, \citenamefont {Frolov},
  \citenamefont {O'Reilly},\ and\ \citenamefont {Fritschel}}]{Gretarsson:07}%
  \BibitemOpen
  \bibfield  {author} {\bibinfo {author} {\bibfnamefont {A.~M.}\ \bibnamefont
  {Gretarsson}}, \bibinfo {author} {\bibfnamefont {E.}~\bibnamefont
  {D'Ambrosio}}, \bibinfo {author} {\bibfnamefont {V.}~\bibnamefont {Frolov}},
  \bibinfo {author} {\bibfnamefont {B.}~\bibnamefont {O'Reilly}}, \ and\
  \bibinfo {author} {\bibfnamefont {P.~K.}\ \bibnamefont {Fritschel}},\ }\href
  {\doibase 10.1364/JOSAB.24.002821} {\bibfield  {journal} {\bibinfo  {journal}
  {J. Opt. Soc. Am. B}\ }\textbf {\bibinfo {volume} {24}},\ \bibinfo {pages}
  {2821} (\bibinfo {year} {2007})}\BibitemShut {NoStop}%
\bibitem [{\citenamefont {Schnabel}(2017)}]{SCHNABEL20171}%
  \BibitemOpen
  \bibfield  {author} {\bibinfo {author} {\bibfnamefont {R.}~\bibnamefont
  {Schnabel}},\ }\href {\doibase https://doi.org/10.1016/j.physrep.2017.04.001}
  {\bibfield  {journal} {\bibinfo  {journal} {Physics Reports}\ }\textbf
  {\bibinfo {volume} {684}},\ \bibinfo {pages} {1 } (\bibinfo {year} {2017})},\
  \bibinfo {note} {squeezed states of light and their applications in laser
  interferometers}\BibitemShut {NoStop}%
\bibitem [{\citenamefont {Zhang}(2003)}]{PhysRevA.67.054302}%
  \BibitemOpen
  \bibfield  {author} {\bibinfo {author} {\bibfnamefont {J.}~\bibnamefont
  {Zhang}},\ }\href {\doibase 10.1103/PhysRevA.67.054302} {\bibfield  {journal}
  {\bibinfo  {journal} {Phys. Rev. A}\ }\textbf {\bibinfo {volume} {67}},\
  \bibinfo {pages} {054302} (\bibinfo {year} {2003})}\BibitemShut {NoStop}%
\bibitem [{\citenamefont {Ma}\ \emph {et~al.}(2017)\citenamefont {Ma},
  \citenamefont {Miao}, \citenamefont {Pang}, \citenamefont {Evans},
  \citenamefont {Zhao}, \citenamefont {Harms}, \citenamefont {Schnabel},\ and\
  \citenamefont {Chen}}]{Ma:2017aa}%
  \BibitemOpen
  \bibfield  {author} {\bibinfo {author} {\bibfnamefont {Y.}~\bibnamefont
  {Ma}}, \bibinfo {author} {\bibfnamefont {H.}~\bibnamefont {Miao}}, \bibinfo
  {author} {\bibfnamefont {B.~H.}\ \bibnamefont {Pang}}, \bibinfo {author}
  {\bibfnamefont {M.}~\bibnamefont {Evans}}, \bibinfo {author} {\bibfnamefont
  {C.}~\bibnamefont {Zhao}}, \bibinfo {author} {\bibfnamefont {J.}~\bibnamefont
  {Harms}}, \bibinfo {author} {\bibfnamefont {R.}~\bibnamefont {Schnabel}}, \
  and\ \bibinfo {author} {\bibfnamefont {Y.}~\bibnamefont {Chen}},\ }\href
  {https://doi.org/10.1038/nphys4118} {\bibfield  {journal} {\bibinfo
  {journal} {Nature Physics}\ }\textbf {\bibinfo {volume} {13}},\ \bibinfo
  {pages} {776 EP } (\bibinfo {year} {2017})}\BibitemShut {NoStop}%
\bibitem [{\citenamefont {Danilishin}\ \emph {et~al.}(2019)\citenamefont
  {Danilishin}, \citenamefont {Khalili},\ and\ \citenamefont
  {Miao}}]{Danilishin2019}%
  \BibitemOpen
  \bibfield  {author} {\bibinfo {author} {\bibfnamefont {S.~L.}\ \bibnamefont
  {Danilishin}}, \bibinfo {author} {\bibfnamefont {F.~Y.}\ \bibnamefont
  {Khalili}}, \ and\ \bibinfo {author} {\bibfnamefont {H.}~\bibnamefont
  {Miao}},\ }\href {\doibase 10.1007/s41114-019-0018-y} {\bibfield  {journal}
  {\bibinfo  {journal} {Living Reviews in Relativity}\ }\textbf {\bibinfo
  {volume} {22}},\ \bibinfo {pages} {2} (\bibinfo {year} {2019})}\BibitemShut
  {NoStop}%
\bibitem [{\citenamefont {Kimble}\ \emph {et~al.}(2001)\citenamefont {Kimble},
  \citenamefont {Levin}, \citenamefont {Matsko}, \citenamefont {Thorne},\ and\
  \citenamefont {Vyatchanin}}]{kimble2001}%
  \BibitemOpen
  \bibfield  {author} {\bibinfo {author} {\bibfnamefont {H.~J.}\ \bibnamefont
  {Kimble}}, \bibinfo {author} {\bibfnamefont {Y.}~\bibnamefont {Levin}},
  \bibinfo {author} {\bibfnamefont {A.~B.}\ \bibnamefont {Matsko}}, \bibinfo
  {author} {\bibfnamefont {K.~S.}\ \bibnamefont {Thorne}}, \ and\ \bibinfo
  {author} {\bibfnamefont {S.~P.}\ \bibnamefont {Vyatchanin}},\ }\href
  {\doibase 10.1103/PhysRevD.65.022002} {\bibfield  {journal} {\bibinfo
  {journal} {Phys. Rev. D}\ }\textbf {\bibinfo {volume} {65}},\ \bibinfo
  {pages} {022002} (\bibinfo {year} {2001})}\BibitemShut {NoStop}%
\bibitem [{\citenamefont {Danilishin}\ and\ \citenamefont
  {Khalili}(2012)}]{SD2012}%
  \BibitemOpen
  \bibfield  {author} {\bibinfo {author} {\bibfnamefont {S.~L.}\ \bibnamefont
  {Danilishin}}\ and\ \bibinfo {author} {\bibfnamefont {F.~Y.}\ \bibnamefont
  {Khalili}},\ }\href {http://www.livingreviews.org/lrr-2012-5} {\bibfield
  {journal} {\bibinfo  {journal} {Living Reviews in Relativity}\ }\textbf
  {\bibinfo {volume} {15}} (\bibinfo {year} {2012})}\BibitemShut {NoStop}%
\bibitem [{\citenamefont {Izumi}\ and\ \citenamefont
  {Sigg}(2016)}]{Izumi_2016}%
  \BibitemOpen
  \bibfield  {author} {\bibinfo {author} {\bibfnamefont {K.}~\bibnamefont
  {Izumi}}\ and\ \bibinfo {author} {\bibfnamefont {D.}~\bibnamefont {Sigg}},\
  }\href {\doibase 10.1088/0264-9381/34/1/015001} {\bibfield  {journal}
  {\bibinfo  {journal} {Classical and Quantum Gravity}\ }\textbf {\bibinfo
  {volume} {34}},\ \bibinfo {pages} {015001} (\bibinfo {year}
  {2016})}\BibitemShut {NoStop}%
\bibitem [{\citenamefont {Tse}\ \emph {et~al.}(2019)\citenamefont {Tse},
  \citenamefont {Yu}, \citenamefont {Kijbunchoo}, \citenamefont
  {Fernandez-Galiana}, \citenamefont {Dupej},\ and\ \citenamefont
  {Barsotti}}]{Tse_SQZ_2019}%
  \BibitemOpen
  \bibfield  {author} {\bibinfo {author} {\bibfnamefont {M.}~\bibnamefont
  {Tse}}, \bibinfo {author} {\bibfnamefont {H.}~\bibnamefont {Yu}}, \bibinfo
  {author} {\bibfnamefont {N.}~\bibnamefont {Kijbunchoo}}, \bibinfo {author}
  {\bibfnamefont {A.}~\bibnamefont {Fernandez-Galiana}}, \bibinfo {author}
  {\bibfnamefont {P.}~\bibnamefont {Dupej}}, \ and\ \bibinfo {author}
  {\bibfnamefont {L.~e.~a.}\ \bibnamefont {Barsotti}},\ }\href {\doibase
  10.1103/PhysRevLett.123.231107} {\bibfield  {journal} {\bibinfo  {journal}
  {Phys. Rev. Lett.}\ }\textbf {\bibinfo {volume} {123}},\ \bibinfo {pages}
  {231107} (\bibinfo {year} {2019})}\BibitemShut {NoStop}%
\bibitem [{\citenamefont {Araj}(2016)}]{Araj_OMC_Scan_1988}%
  \BibitemOpen
  \bibfield  {author} {\bibinfo {author} {\bibfnamefont {K.}~\bibnamefont
  {Araj}},\ }\href
  {https://alog.ligo-la.caltech.edu/aLOG/index.php?callRep=25070} {\enquote
  {\bibinfo {title} {Advanced ligo livingston logbook report 25070},}\ }
  (\bibinfo {year} {2016})\BibitemShut {NoStop}%
\bibitem [{\citenamefont {Oelker}(2016)}]{Oelker_THESIS_2016}%
  \BibitemOpen
  \bibfield  {author} {\bibinfo {author} {\bibfnamefont {E.}~\bibnamefont
  {Oelker}},\ }\emph {\bibinfo {title} {Squeezed States for Advanced
  Gravitational Wave Detectors}},\ \href
  {https://dcc.ligo.org/DocDB/0127/P1600245/001/main.pdf} {Ph.D. thesis},\
  \bibinfo  {school} {MIT} (\bibinfo {year} {2016})\BibitemShut {NoStop}%
\bibitem [{\citenamefont {Purdue}\ and\ \citenamefont
  {Chen}(2002)}]{PhysRevD.66.122004}%
  \BibitemOpen
  \bibfield  {author} {\bibinfo {author} {\bibfnamefont {P.}~\bibnamefont
  {Purdue}}\ and\ \bibinfo {author} {\bibfnamefont {Y.}~\bibnamefont {Chen}},\
  }\href {\doibase 10.1103/PhysRevD.66.122004} {\bibfield  {journal} {\bibinfo
  {journal} {Phys. Rev. D}\ }\textbf {\bibinfo {volume} {66}},\ \bibinfo
  {pages} {122004} (\bibinfo {year} {2002})}\BibitemShut {NoStop}%
\bibitem [{\citenamefont {Walls}\ and\ \citenamefont
  {Milburn}(2007)}]{walls2007quantum}%
  \BibitemOpen
  \bibfield  {author} {\bibinfo {author} {\bibfnamefont {D.~F.}\ \bibnamefont
  {Walls}}\ and\ \bibinfo {author} {\bibfnamefont {G.~J.}\ \bibnamefont
  {Milburn}},\ }\href@noop {} {\emph {\bibinfo {title} {Quantum optics}}}\
  (\bibinfo  {publisher} {Springer Science \& Business Media},\ \bibinfo {year}
  {2007})\BibitemShut {NoStop}%
\end{thebibliography}%
\end{document}